\documentclass[12pt]{article}
\usepackage[utf8]{inputenc}

\usepackage{amsmath}
\usepackage{amssymb}
\begin{document}

\newcommand{\br}{|\kern-.25em|\kern-.25em|}

\renewcommand{\theequation}{\thesection.\arabic{equation}}
\newtheorem{theorem}{Theorem}[section]

\newtheorem{definition}[theorem]{Definition}
\newtheorem{lemma}[theorem]{Lemma}
\newtheorem{example}[theorem]{Example}
\newtheorem{remark}[theorem]{Remark}
\newtheorem{remarks}[theorem]{Remarks}
\newtheorem{cor}[theorem]{Corollary}
\newtheorem{pro}[theorem]{Proposition}

\begin{titlepage}
\begin{center}
{\Large\bf Space--time statistical solutions\\\vspace{0.5cm}
for an inhomogeneous chain of harmonic oscillators}\\
\vspace{1cm}
{\large T.V. Dudnikova}
\\
\vspace{0.5cm}
{\small\it Keldysh Institute of Applied Mathematics\\
Russian Academy of Science\\
 Moscow  125047 Russia}\\
~\\
e-mail: tdudnikov@mail.ru
\end{center}
 \vspace{2cm}

  \begin{abstract}
We consider an one-dimensional inhomogeneous harmonic chain consisting
of two different semi-infinite chains of harmonic  oscillators.
We study the Cauchy problem with random initial data.
Under some restrictions on the interaction between the oscillators of the chain
and on the distribution of the initial data,
we prove the convergence of space-time statistical solutions to  a Gaussian measure.
\medskip

\medskip

 \end{abstract}
\end{titlepage}

\section{Introduction}
We consider an infinite one-dimensional harmonic chain of particles
having nearest-neighbor interactions and unit mass.
We assume that the particles located at points $x=1,2,\ldots$
have the same interaction force constants $\nu_+>0$,
and the same external harmonic forces with constants  $\kappa_+\ge0$ act on them.
The particles located at points $x=-1,-2,\ldots$
 have  constants  $\nu_->0$ and $\kappa_-\ge0$, respectively.
In addition,
an external force with a constant $\kappa_0\ge0$ acts on the particle located at the origin,
and $\kappa_0\not=\kappa_\pm$, in general.
Therefore, the displacement of the particle located at a point $x\in\mathbb{Z}$
from its equilibrium position obeys the following equations:
\begin{eqnarray}\label{1}
 \left\{
 \begin{array}{ll} \ddot u(x,t)=(\nu^2_+\Delta_L-\kappa_+^2) u(x,t),& x\ge1,\quad t>0,
\\
\ddot u(0,t)=\nu^2_+(u(1,t)-u(0,t))+\nu^2_{-}(u(-1,t)-u(0,t))
-\kappa_0^2u(0,t),&t>0,\\
 \ddot u(x,t)=(\nu^2_-\Delta_L-\kappa_-^2) u(x,t),&x\le-1,\quad t>0.
\end{array}\right.\end{eqnarray}
Here $u(x,t)\in\mathbb{R}$,  $\Delta_L$ denotes the second derivative on
$\mathbb{Z}=\{0,\pm1,\pm2,\dots\}$:
$$
\Delta_L u(x)=u(x+1)-2u(x)+u(x-1),\qquad x\in\mathbb{Z}.
$$
For system (\ref{1}), we study the Cauchy problem with the initial data
 \begin{equation}\label{4}
 u(x,0) = u_{0}(x),\qquad \dot u(x,0) = v_{0}(x), \qquad x\in\mathbb{Z}.
\end{equation}
Formally, this system is Hamiltonian with the Hamiltonian functional of the form
  \begin{gather*}
 {\rm H}(u,\dot u)=
 {\rm H}_+(u,\dot u)+{\rm H}_-(u,\dot u)+{\rm H}_0(u,\dot u),\\
 {\rm H}_\pm(u,\dot u):=
\frac12\sum\limits_{\pm x\ge1}\Big(
|\dot u(x,t)|^2+\nu_\pm^2|u(x\pm1,t)-u(x,t)|^2+\kappa_\pm^2|u(x,t)|^2\Big),\\
{\rm H}_0(u,\dot u):=\frac12\Big(
|\dot u(0,t)|^2+\sum_\pm\nu_\pm^2|u(\pm1,t)-u(0,t)|^2+\kappa_0^2|u(0,t)|^2\Big).
\end{gather*}
We consider two cases of equations~(\ref{1}).
In the first case, we assume that the harmonic chain is homogeneous
and $\kappa_\pm,\kappa_0>0$, i.e.,
\begin{equation}\label{0.1}
\nu_\pm=:\nu>0\quad \mbox{and }\,\,\kappa_\pm=\kappa_0=:\kappa>0.
\end{equation}
In the second one, we impose  condition~{\bf C}
on the coefficients $\nu_\pm>0$ and $\kappa_0,\kappa_\pm\ge0$.
To state this condition, we introduce the following notation.
For simplicity, we assume that $\kappa_-\le \kappa_+$.
Put $a_\pm:=\sqrt{4\nu^2_\pm+\kappa_\pm^2}$ and
\begin{gather*}
   K_\pm(\omega):=\frac12\left(\kappa_-^2+\kappa_+^2\right)+\frac12\sqrt{\omega^2-\kappa_\pm^2}\sqrt{\omega^2-a_\pm^2}
 \quad \mbox{for }\,\, \omega\in\mathbb{R}:\,\,\,|\omega|\ge a_\pm;   \\
 K_0(\omega):=\frac12\left(\kappa_-^2+\kappa_+^2\right)-\frac12\sqrt{\kappa_+^2-\omega^2}\sqrt{a_+^2-\omega^2}
 \quad\mbox{for }\,\,  \omega\in\mathbb{R}:\,\,|\omega|\le \kappa_+\,\,(\mbox{if }\,\kappa_+>0).
\end{gather*}

{\bf Condition~C.}
For different values of $\kappa_\pm$ and $\nu_\pm$, the constant $\kappa_0$
satisfies the following restrictions:
\begin{align*}
&\kappa^2_0< K_+(a_-),\quad\mbox{if }\,\,\,a_-\ge a_+;\qquad \kappa^2_0<K_-(a_+),\quad\mbox{if }\,\,\,a_+\ge a_-;\\
&\kappa^2_0> K_0(\kappa_-),\quad\mbox{if }\,\,\,\kappa_-\not=0;\\
&\kappa^2_0> K_-(\kappa_+)\,\,\,\mbox{or }\,\, \kappa^2_0< K_0(a_-),\quad\mbox{if }\,\,a_-\le\kappa_+;\\
&\kappa_0\not=0,\quad\mbox{if }\,\, \kappa_-=\kappa_+=0.
\end{align*}
Note that if $\kappa_+=\kappa_-$, then  condition~$\mathbf{C}$  implies that
$$
\kappa_0^2\in\left(\kappa_-^2,\kappa_-^2
+2\max\{\nu_-,\nu_+\}\sqrt{|\nu_-^2-\nu_+^2|}\right)\quad \mbox{and }\,\,
\nu_-\not=\nu_+.
$$
Thus, condition~$\mathbf{C}$ excludes the case when
two semi-infinite parts of the chain are identical, i.e.,
when $\kappa_+=\kappa_-$ and $\nu_+=\nu_-$.
\medskip

 We assume that the initial data $Y_0$
belong to the phase space ${\cal H}_\alpha$, $\alpha\in\mathbb{R}$,
defined below.
 \begin{definition}   \label{d1.1}
 (i)
 $\ell^2_{\alpha}\equiv\ell^2_{\alpha}(\mathbb{Z})$,
$\alpha\in\mathbb{R}$,  is the  Hilbert space of real-valued sequences $u(x)$, $x\in\mathbb{Z}$,  with the norm
$$ \Vert u\Vert_{\alpha}
=\Big(\sum\limits_{x\in\mathbb{Z}}\langle x\rangle^{2\alpha}u^2(x)\Big)^{1/2}<\infty,\quad
\langle x\rangle:=(1+x^2)^{1/2}.
$$
Below we use also the notation $\ell^2\equiv \ell^2_0$.

 (ii)
  $\mathcal{H}_\alpha=\ell^2_{\alpha}\times \ell^2_{\alpha}$ is the  Hilbert space
  of pairs $Y=(u(x),v(x))$  of  real-valued sequences $u(x)$ and $v(x)$ endowed with  the norm
$$
 \Vert Y\Vert^2_{\alpha} =\Vert u\Vert^2_{\alpha}+\Vert v\Vert^2_{\alpha} <\infty.
$$
 (iii) Write $\mathfrak{C}^k_{\alpha}=C^k(\mathbb{R};\ell^2_{\alpha})$, $k=0,1$, $\alpha\in\mathbb{R}$.
Introduce the seminorms in $\mathfrak{C}^k_{\alpha}$ by the rule
\begin{equation}\label{1.3}
\br u(\cdot,\cdot)\br^2_{\alpha,k,T}=\max_{|t|\le T}\sum\limits_{r=0}^k
\Vert \partial _t^r u(\cdot,t)\Vert^2_{\alpha},\qquad T>0.
\end{equation}
(iv) Denote by $R$ the operator $R:\mathcal{H}_{\alpha}\to\mathfrak{C}^1_{\alpha}$ such that
\begin{equation}\label{R}
(RY_0)(x,t)=u(x,t),
\end{equation}
where $u(x,t)$ is the solution to problem~(\ref{1})--(\ref{4}) with the initial data $Y_0=(u_0,v_0)$.
   \end{definition}

Below we assume that $\alpha<-3/2$ if condition~{\bf C} holds and
$\alpha<-1/2$ if condition~(\ref{0.1}) holds.
We suppose that the initial date $Y_0$ is a random function.
Denote by $\mu_0$  a Borel probability measure on $\mathcal{H}_{\alpha}$ giving the distribution of $Y_0$.
\begin{definition}\label{def1.2}
Introduce a Borel probability measure $P$ on the space $\mathfrak{C}^1_{\alpha}$ by the rule
$$
P(\omega)=\mu_0(R^{-1}\omega)  \quad \mbox{for any Borel set }\,\,\omega\in{\cal B}(\mathfrak{C}^1_\alpha).
$$
Here and below  ${\cal B}(X)$  denotes the $\sigma$-algebra of Borel  sets of a topological space
$X$.
The measure $P$ is called a space-time statistical solution to problem (\ref{1})--(\ref{4})
corresponding to the initial measure $\mu_0$.
Denote by $\{P_\tau, \tau\in\mathbb{R}\}$ the following family of measures
$$
P_\tau(\omega)=P(S_\tau^{-1}\omega)\quad
\mbox{for any }\,\, \omega\in{\cal B}({\mathfrak C}^1_\alpha),\quad \tau\in\mathbb{R}.
$$
Here $S_\tau$ denotes the shift operator in time,
\begin{equation}\label{1.5}
S_\tau(u(x,t))=u(x,t+\tau), \qquad \tau\in\mathbb{R}.
\end{equation}
\end{definition}

The main goal of the paper is to prove that the measures $P_\tau$ weakly converge as $\tau\to\infty$
to  a limit on the space $\mathfrak{C}^0_\alpha$,
\begin{equation}\label{1.8i}
P_\tau\rightharpoondown P_\infty,\quad \tau\to\infty.
\end{equation}
This means the convergence of the integrals
$$
 \int_{\mathfrak{C}^0_\alpha} f(u)P_\tau(du)\rightarrow
  \int_{\mathfrak{C}^0_\alpha} f(u)P_\infty(du)
 \quad \mbox{as }\,\,\, \tau\to \infty
$$
 for any bounded continuous functional $f$  on $\mathfrak{C}^0_\alpha$.
Furthermore, the limit measure $P_\infty$ is a Gaussian measure on the space
 $\mathfrak{C}^1_{\alpha}$ supported by the solutions to problem~(\ref{1}).
 Thus, the convergence~(\ref{1.8i}) can be considered as an analog
 of the central limit theorem for a class of solutions to the equations~(\ref{1}).
The proof of convergence~(\ref{1.8i}) is based on the results of \cite{D20}
and used the technique of  \cite{KR, VF}.
 Also, we check  that the group $S_\tau$ is mixing w.r.t.
 the measure $P_\infty$, i.e.,
  for any $f,g\in L^{2}(\mathfrak{C}^1_\alpha,P_\infty)$,
\begin{equation}\label{2.33}
\lim_{\tau\to\infty} \int
 f(S_{\tau} u)g(u)\,P_\infty(du)
 =\int f(u)P_\infty(du)\int g(u)P_\infty(du).
\end{equation}

For models described by partial differential equations,
the long-time behavior of space-time statistical solutions
was studied by Komech and Ratanov~\cite{KR}
for wave equations and  Ratanov~\cite{R91} for parabolic equations.
For Klein--Gordon equations, the result was obtained in \cite{D7}.
The time evolution and
ergodic properties of infinite harmonic crystals were studied
by Lanford, Lebowitz~\cite{LL} and by van~Hemmen~\cite{Hem}.
For the one-dimensional chains of harmonic oscillators,
the  behavior of {\em statistical} solutions $\mu_t:=[U(t)]^*\mu_0$ as $t\to\infty$,
where $U(t)$ stands for the solving operator of  problem~(\ref{1})--(\ref{4}),
was investigated in \cite{D20}.
In this paper, we extend these results to the {\em space-time statistical}  solutions of problem~(\ref{1}).

\setcounter{equation}{0}
\section{Main results}

Introduce the notation
\begin{equation*} 
Y_0(x)=(Y_0^0(x),Y_0^1(x))\equiv (u_0(x),v_0(x)),\quad
Y(t)=(Y^0(t),Y^1(t))\equiv (u(\cdot,t),\dot u(\cdot,t)).
\end{equation*}
\begin{theorem} (see \cite[Theorem~2.1]{D18})
Let $\kappa_\pm,\kappa_0\ge0$, $\nu_\pm>0$
and $Y_0\in\mathcal{H}_{\alpha}$, $\alpha\in\mathbb{R}$.
Then  the Cauchy problem~(\ref{1})--(\ref{4}) has a unique solution
 $Y(t)\in C(\mathbb{R},\mathcal{H}_{\alpha})$.
 The operator
$U(t):Y_0\to Y(t)$ is continuous in $\mathcal{H}_{\alpha}$. Furthermore,
there exist constants $C,B<\infty$ such that
\begin{equation}\label{2.5}
 \Vert U(t)Y_0\Vert_{\alpha}\le Ce^{B|t|}\Vert Y_0\Vert_{\alpha}\,,\qquad t\in\mathbb{R}.
\end{equation}
\end{theorem}
\begin{cor}
It follows from (\ref{2.5}) that for any $Y_0\in\mathcal{H}_{\alpha}$,
$$
\br RY_0\br_{\alpha,1,T}\le C(T)\Vert Y_0\Vert_{\alpha}, \quad
\forall\, T>0,
$$
where the operator $R$ is defined in (\ref{R}).
\end{cor}

Below we assume that
$\alpha<-1/2$ if condition~(\ref{0.1}) holds
 and $\alpha<-3/2$ if condition~{\bf C} holds.

\subsection{Conditions on the initial measure}
\begin{definition}
(i) A measure $\mu$ is called  translation invariant (or space homogeneous) if
$\mu(\mathcal{S}_h B)= \mu(B)$ for any $B\in{\cal B}(\mathcal{H}_{\alpha})$ and
$h\in\mathbb{Z}$, where $\mathcal{S}_h Y(x)= Y(x+h)$, $x\in\mathbb{Z}$.

(ii)
For a probability  measure $\mu$ on  $\mathcal{H}_{\alpha}$,
we denote by $\hat\mu$ its characteristic functional (Fourier transform),
$$
\hat\mu(\Psi)\equiv\int\exp(i\langle Y,\Psi\rangle)\,\mu(dY), \qquad \Psi\in{\cal S}.
$$
Here $\Psi=(\Psi^0,\Psi^1)\in{\cal S}:=S\oplus S$,
$S:=S(\mathbb{Z})$,
where $S(\mathbb{Z})$ denotes a space of real  quickly decreasing  sequences,
$$
\langle Y,\Psi \rangle
=\sum\limits_{i=0,1}\sum\limits_{x\in\mathbb{Z}} Y^i(x) \Psi^i(x),
\qquad Y=(Y^0,Y^1), \quad \Psi=(\Psi^0,\Psi^1).
$$
Below we use also the notation
 $ \langle Y,\Psi \rangle_\pm
:=\sum\limits_{i=0,1}\sum\limits_{x\in\mathbb{Z}_\pm} Y^i(x)\Psi^i(x)$,
$\mathbb{Z}_{\pm}:=\{x\in\mathbb{Z}:\pm x\ge0\}$.

(iii)
A measure $\mu$ is called Gaussian (of zero mean) if
its characteristic functional has the form
$\hat{\mu}(\Psi)=  \exp\{-{\cal Q}(\Psi,\Psi)/2\}$,
where ${\cal Q}$ is a real-valued nonnegative quadratic form in ${\cal S}$.
\end{definition}

We assume that the initial data $Y_0(x)$
in (\ref{4}) is a measurable random function with values in
 $(\mathcal{H}_{\alpha},\,{\cal B}(\mathcal{H}_{\alpha}))$.
Recall that $\mu_0$ is a Borel probability measure on  $\mathcal{H}_{\alpha}$
which is the distribution of $Y_0$. Let $\mathbb{E}$
stand for the mathematical expectation w.r.t. this measure.
Denote by $Q_0(x,y)=\left(Q_0^{ij}(x,y)\right)_{i,j=0,1}$
the correlation matrix of the measure $\mu_0$, where
\begin{equation*}
Q^{ij}_0(x,y):=\mathbb{E}\left(Y^i_0(x) Y^j_0(y) \right)\equiv
\int Y^i_0(x)Y^j_0(y)\,\mu_0(dY_0),\qquad x,y\in\mathbb{Z},\qquad i,j=0,1,
\end{equation*}
and by ${\cal Q}_0(\Psi,\Psi)$  a  real-valued quadratic form
on ${\cal S}$ with the matrix kernel $Q_0(x,y)$.
\medskip

We impose conditions {\bf S1}--{\bf S4} on the initial measure $\mu_0$.
\begin{itemize}
\item[{\bf S1}]
 $\mu_0$ has zero mean value, i.e.,
$\mathbb{E} \left(Y_0(x)\right)= 0$,
$x\in\mathbb{Z}$.

\item[{\bf S2}]  The correlation functions $Q^{ij}_0(x,y)$ satisfy the  bound
\begin{equation}
\label{Qh}
|Q^{ij}_0(x,y)|\le h(|x-y|),
\end{equation}
where $h$ is a nonnegative bounded function and
$h(r)\in L^1(0,+\infty)$.

\item[{\bf  S3}]
The correlation matrix  $Q_0(x,y)$ satisfies the
following condition
 \begin{eqnarray}\label{1.7'}
Q_0(x+y, y)\to\left\{\begin{array}{lll}
q_-(x) &\mbox{ as}& y\to-\infty\\
q_+(x)&\mbox{ as}& y\to+\infty
\end{array} \right|\quad x\in\mathbb{Z}.
\end{eqnarray}
Here  $q_{\pm}(x)=\left(q^{ij}_{\pm}(x)\right)_{i,j=0,1}$  stand for
correlation matrices of some translation invariant measures
 $\mu_{\pm}$ with zero mean in $\mathcal{H}_{\alpha}$.
 \end{itemize}
\begin{definition}
Let ${\cal A}$ be an interval in $\mathbb{Z}$.
Denote by $\sigma({\cal A})$
a $\sigma$-algebra in  $\mathcal{H}_{\alpha}$
generated by the initial data $Y_0(x)$ with $x\in{\cal A}$.
Introduce the Ibragimov mixing coefficient of the measure $\mu_0$  by the rule
\begin{equation*}
\varphi(r)\equiv
\sup \frac{| \mu_0(A\cap B) - \mu_0(A)\mu_0(B)|}{ \mu_0(B)}\,.
\end{equation*}
Here the supremum is taken over all sets
$A\in\sigma({\cal A})$, $B\in\sigma({\cal B})$ with $\mu_0(B)>0$, and
all intervals ${\cal A},{\cal B}\subset \mathbb{Z}$ with distance $\rho({\cal A},{\cal B})\ge r$.
The measure $\mu_0$  satisfies   Ibragimov's strong uniform mixing condition if $\varphi(r)\to 0$ as $r\to\infty$
(cf. \cite[Definition 17.2.2]{IL}).
\end{definition}
\begin{itemize}
\item[{\bf S4}]
$\mu_0$  has a finite ``mean energy density'', i.e.,
$ \sup\limits_{x\in\mathbb{Z}}\mathbb{E} |Y_0(x)|^2 \le e_0<\infty$.
Moreover,  $\mu_0$ satisfies Ibragimov's strong uniform  mixing condition, and $\varphi^{1/2}(r)\in L^1(0,+\infty)$.
\end{itemize}
\begin{lemma}\label{rem2.5}
(i) Condition  {\bf S2} implies that for any
 $\Phi,\Psi\in \mathcal{H}_0$,
\begin{equation}\label{c4.1}
|{\cal Q}_0(\Phi,\Psi)|\equiv
|\langle Q_0(x,y),\Phi(x)\otimes\Psi(y)\rangle|\le
C\Vert\Phi\Vert_{0}\Vert\Psi\Vert_{0}.
\end{equation}
This follows from the bound~(\ref{Qh}) applying either
 the Shur test (see, e.g., \cite[p.223]{Ni})
 or Young's inequality (see, e.g., \cite[Theorem~0.3.1]{So}).

(ii) It follows from conditions {\bf S1}--{\bf S3} that $q^{ij}_\pm\in\ell^1$, $i,j=0,1$.
Hence, $\hat q^{ij}_\pm\in C(\mathbb{T})$.
Assertions~(i) and (ii) are proved in \cite[Lemma~5.1]{D19}.

(iii) Conditions {\bf S1} and {\bf S4} imply the bound~(\ref{Qh})
with the function $h(r)=Ce_0 \varphi^{1/2}(r)$.
This follows from \cite[Lemma~17.2.3]{IL}.

(iv) The correlation functions $Q_0^{ij}$ have the property:
$Q_0^{ij}(x,y)= Q_0^{ji}(y,x)$, $i,j=0,1$.
Then, the correlation functions $q_\pm^{ij}$ from condition~{\bf S3}
satisfy the  relation
\begin{equation}\label{2.7}
q_\pm^{ii}(-x)=q_\pm^{ii}(x),\qquad q_\pm^{10}(x)=q_\pm^{01}(-x),
\qquad x\in\mathbb{Z}.
\end{equation}
\end{lemma}

\subsection{The convergence of space--time statistical solutions}
Denote by $\mathcal{P}$  a space of real-valued functions $v(x,t)$
which are infinite differentiable in $t$ and  quickly decrease in $t$ and $x$,
$$
\sup\limits_{t\in\mathbb{R}}\sup\limits_{x\in\mathbb{Z}}
\langle x\rangle^M\langle t\rangle^N|\partial_t^rv(x,t)|\le C<\infty
\quad\mbox{for any }\, M,N\quad \mbox{and }\,r\ge0.
$$
Let $[\cdot,\cdot]$ stand for the inner product in $L^2(\mathbb{R};\ell^2)$ (or in its extensions),
$$
[u_1,u_2]=\sum\limits_{x\in\mathbb{Z}}\,\, \int\limits_{-\infty}^{+\infty}u_1(x,t) u_2(x,t)\,dt.
$$
\begin{definition}\label{def2.6}
Denote by $Q^P_\tau(x_1,x_2,t_1,t_2)$, $x_1,x_2\in\mathbb{Z}$, $t_1,t_2\in\mathbb{R}$,
the correlation functions of the measures $P_\tau$, $\tau\in\mathbb{R}$, introduced in Definition~\ref{def1.2},
i.e., for any $v_1,v_2\in\mathcal{P}$,
\begin{equation*}
\begin{split}
\mathcal{Q}^P_\tau(v_1,v_2)&:=[Q^P_\tau,v_1\otimes v_2]=
\int[u,v_1][u,v_2]\,P_\tau(du)\\
&=
\sum\limits_{x_1,x_2\in\mathbb{Z}}\,\,\int\limits_{-\infty}^{+\infty}dt_1\int\limits_{-\infty}^{+\infty}
Q^P_\tau(x_1,x_2,t_1,t_2)v_1(x_1,t_1) v_2(x_2,t_2)\,dt_2,
\qquad\tau\in\mathbb{R}.
\end{split}
\end{equation*}
\end{definition}

The main result of the paper is the following theorem.
\begin{theorem}\label{tA}
   Let   $\alpha<-3/2$ and condition~{\bf C} hold.
  Then the following assertions hold.

(i) Let conditions {\bf S1}--{\bf S3} be fulfilled.
Then the correlation functions of $P_\tau$ converge to a limit as $\tau\to\infty$.
 Moreover,
 for any $v_1,v_2\in\mathcal{P}$,
\begin{equation}\label{i}
 \mathcal{Q}_\tau^P(v_1,v_2)\to  \mathcal{Q}_\infty^P(v_1,v_2)\qquad \mbox{as }\,\,\, \tau\to\infty,
\end{equation}
where
 \begin{equation}\label{2.9}
 \mathcal{Q}_\infty^P(v_1,v_2)=\mathcal{Q}_\infty^{P,\nu}(T\Omega'\vec{v}_1,T\Omega'\vec{v}_2),
 \end{equation}
 $\vec{v}_i:=(v_i,0)$,
 the quadratic form $\mathcal{Q}_\infty^{P,\nu}$ is defined in (\ref{3.14}) below,
 the operators $\Omega'$ and $T$  are defined in (\ref{operatorOmega'}) and (\ref{operatorT}),
 respectively.

(ii) Let conditions {\bf S1}, {\bf S3}, and {\bf S4} be fulfilled.
Then the convergence~(\ref{1.8i}) holds.
 The limit measure $P_\infty$ is a Gaussian measure on the space
 $\mathfrak{C}^1_{\alpha}$ supported by the solutions to problem~(\ref{1}).

 (iii)
 The measure $P_\infty$ is invariant w.r.t. the shifts in time,
  and the convergence~(\ref{2.33}) holds.
\end{theorem}
{\bf Remark}.
\emph{If the initial measure $\mu_0$ is Gaussian, then convergence~(\ref{1.8i})
follows from convergence~(\ref{i}).
Furthermore, the weak convergence of the measures $P_\tau$
doesn't imply, in general, the convergence of their correlation matrices.
Therefore, the last fact we prove separately.}
\smallskip

Theorem~\ref{tA} is proved in Section~\ref{sec2.5}.
In Appendix, we consider the homogeneous case~(\ref{0.1}) and prove the similar results.
\begin{theorem}\label{tB}
   Let   $\alpha<-1/2$ and condition~(\ref{0.1}) hold.
  Then all assertions of Theorem~\ref{tA} remain true with
  the limiting correlation function $Q^P_\infty(x_1,x_2,t_1,t_2)$ of the following form
 $$
 Q_\infty^P(x_1,x_2,t_1,t_2)=q^P_\infty(x_1-x_2,t_1-t_2).
 $$
The Fourier transform  of $q^P_\infty(x,t)$ w.r.t. variable $x$ ($x\to\theta$) is
of the form
\begin{equation}\label{2.21}
\hat q^P_\infty(\theta,t)=\cos(\phi(\theta)t)\,\hat q_\infty^{00}(\theta)-
\sin(\phi(\theta)t)\,\phi^{-1}(\theta)\,\hat q_\infty^{01}(\theta),
\end{equation}
where $\phi(\theta):=\sqrt{\nu^2(2-2\cos\theta)+\kappa^2}$, and
\begin{equation}\label{2.8}
\hat q_\infty^{ij}(\theta)=\hat q_{\infty,+}^{ij}(\theta)+\hat q_{\infty,-}^{ij}(\theta),
\quad i,j=0,1,
\end{equation}
with $\hat q^{ij}_{\infty,\pm}(\theta)$
 defined similarly to (\ref{qinfty}) but with $\phi(\theta)$ instead of
$\phi_{\pm}(\theta)$.
\end{theorem}

\setcounter{equation}{0}
\section{Proof: Inhomogeneous case of the chain}\label{sec2.5}
We divide the proof of Theorem~\ref{tA}  into two steps:

{\bf Step 1}: Instead of problem~(\ref{1})--(\ref{4}) we first study a simpler ``unperturbed''
problem~(\ref{a.1}) with zero condition at origin and prove the results
similar to Theorem~\ref{tA}, see Section~\ref{sec3.1}.

{\bf Step 2}: In Section~\ref{sec3.2}, we introduce a ``wave'' operator $\Omega$,
which allows us to reduce the ``perturbed'' problem~(\ref{1})--(\ref{4})
to the problem~(\ref{a.1}).

\subsection{Unperturbed problem}\label{sec3.1}

Consider the following problem
\begin{equation}\label{a.1}
\left\{\begin{array}{ll}
 \ddot z(x,t)=(\nu_\pm^2\Delta_L-\kappa_\pm^2)z(x,t),&\,\, \pm x\ge1,\quad t>0,
\\
z(0,t)=0,&\,\, t\ge0,\\
z(x,0)=u_0(x),\quad \dot z(x,0)=v_0(x),&\,\, x\not=0.
\end{array}
\right.
\end{equation}
 \begin{lemma} (see \cite[Lemma~2.1]{D20})
Let $\alpha\in\mathbb{R}$. Then
for any  $Y_0\equiv(u_0,v_0) \in {\cal H}_{\alpha}$ there exists a unique solution
$Z(t)\equiv(z(\cdot,t),\dot z(\cdot,t))\in C(\mathbb{R}, {\cal H}_{\alpha})$
 to problem~(\ref{a.1}).
Furthermore, the operator $U_0(t):Y_0\mapsto Z(t)$ is continuous in ${\cal H}_{\alpha}$,
and $ \Vert U_0(t)Y_0\Vert_{\alpha}\le Ce^{B|t|}\Vert Y_0\Vert_{\alpha}$, $t\in\mathbb{R}$.
\end{lemma}

The solution to problem~(\ref{a.1}) consists of two solutions to the initial--boundary
value problems in $\mathbb{Z}_+$ and $\mathbb{Z}_-$ with zero boundary condition at $x=0$.
Therefore, the solution to (\ref{a.1}) has a form
\begin{equation}\label{3.2'}
(U_0(t)Y_0)^i(x)=\left\{
\begin{array}{lll}
\sum\limits_{j=0,1}\sum\limits_{y\in\mathbb{Z}_+}
 G^{ij}_{t,+}(x,y) Y_{0}^j(y) &\,\, \mbox{for }& x\in\mathbb{Z}_{+},\\
 \sum\limits_{j=0,1}\sum\limits_{y\in\mathbb{Z}_-}
 G^{ij}_{t,-}(x,y) Y_{0}^j(y)&\,\, \mbox{for }& x\in\mathbb{Z}_{-},
\end{array}
\right.
\end{equation}
where
$Y^0_0(x)\equiv u_0(x)$, $Y_0^1(x)\equiv v_0(x)$, and
the Green function $ G_{t,\pm}(x,y)=( G^{ij}_{t,\pm}(x,y))_{i,j=0}^1$ is a
matrix-valued function with the entries of the form
\begin{gather}
\label{3.2}
G^{ij}_{t,\pm}(x,y):=
{\cal G}^{ij}_{t,\pm}(x-y)-{\cal G}^{ij}_{t,\pm}(x+y),
\quad x,y\in\mathbb{Z}_\pm, \quad
 {\cal G}^{ij}_{t,\pm}(x)\equiv\frac1{2\pi}\int\limits_{\mathbb{T}}
e^{-ix \theta} \hat{\cal G}^{ij}_{t,\pm}(\theta)\,d\theta,\\
\label{hatcalG}
\left(\hat{\cal G}_{t,\pm}^{ij}(\theta)\right)_{i,j=0}^1
=\left(\begin{array}{ll} \cos\left(\phi_\pm(\theta)t\right)&
\sin\left(\phi_\pm(\theta)t\right)/\phi_\pm(\theta)\\
-\phi_\pm(\theta)\sin\left(\phi_\pm(\theta)t\right)& \cos\left(\phi_\pm(\theta)t\right)
\end{array}\right),\\
\label{phi}
\phi_\pm(\theta)=\sqrt{\nu_\pm^2(2-2\cos\theta)+\kappa_\pm^2}.
\end{gather}
In particular,
$\phi_\pm(\theta)=2\nu_\pm|\sin(\theta/2)|$ if $\kappa_\pm=0$.
Note that $G^{ij}_{t,\pm}(0,y)\equiv0$, since
${\cal G}_{t,\pm}^{ij}(-x)={\cal G}_{t,\pm}^{ij}(x)$.
\begin{definition}\label{mu0}
Introduce a measure
 $\nu_0=\mu_0\{Y_0\in{\cal H}_{\alpha}:\, Y_0(0)=0\}$.
 Denote by $\nu_{t}$, $t\in\mathbb{R}$, a Borel probability measure on
${\cal H}_{\alpha}$ giving the distribution of the solution $U_0(t)Y_0$
to problem~(\ref{a.1}), i.e.,
$\nu_{t}(B)=\nu_0(U_0(-t)B)$ for any $B\in{\cal B}({\cal H}_{\alpha})$.
The correlation matrix of  $\nu_t$ is denoted as
\begin{equation*}
Q^{\nu}_t(x,y)=\left(Q^{\nu,ij}_t(x,y)\right)_{i,j=0,1},\quad
Q^{\nu,ij}_t(x,y):=\int Y^i(x)Y^j(y)\,\nu_t(dY),\quad x,y\in\mathbb{Z},\quad t\in\mathbb{R}.
\end{equation*}
\end{definition}

The correlation matrix $Q_t^\nu$ has the following property.
\begin{lemma}\label{l3.3}
Let condition {\bf S2} hold.  Then
\begin{equation}
 \label{Qnu}
\sup\limits_{t\in\mathbb{R}}|Q^\nu_t(x,y)|\le \sqrt{C_1+C_2|x|}\sqrt{C_1+C_2|y|},\quad x,y\in\mathbb{Z},
\end{equation}
where the constants $C_1$ and $C_2$ do not depend on $x,y$,
and  $C_2=0$ if $\kappa_-\kappa_+\not=0$.
 \end{lemma}
 {\bf Proof}\,
We check (\ref{Qnu}) only for $x,y\in\mathbb{Z}_+$. For another values of $x,y$
the proof is similar.
 Using Definition~\ref{mu0} and representation~(\ref{3.2'}), we obtain that for $x,y\in\mathbb{Z}_+$,
   $t\in\mathbb{R}$, $i,j=0,1$,
 \begin{align*}
 Q^{\nu,ij}_t(x,y)&=\int \left(U_0(t)Y_0\right)^i(x)\left(U_0(t)Y_0\right)^j(y)\,\nu_0(dY_0)\\
 &=\sum\limits_{k,l=0,1}
 \sum\limits_{x',y'\in\mathbb{Z}_+}
 G_{t,+}^{ik}(x,x')Q^{\nu,kl}_0(x',y') G_{t,+}^{jl}(y,y')=
 \langle Q^{\nu}_0(\cdot,\cdot),\Phi^{i}_x(\cdot,t)\otimes\Phi^{j}_y(\cdot,t)\rangle_+,
 \end{align*}
 where $\Phi^{i}_x(x',t):= \left(G_{t,+}^{i0}(x,x'),G_{t,+}^{i1}(x,x')\right)$.
Hence, applying (\ref{c4.1}), one obtains
 \begin{equation*}
  \left|Q^{\nu,ij}_t(x,y)\right|\le
  C\Vert\Phi^{i}_x(\cdot,t)\Vert_0\Vert\Phi^{j}_y(\cdot,t)\Vert_0,
     \end{equation*}
where the constant $C$ does not depend on $x,y,t$.
On the other hand, the Parseval identity and (\ref{3.2}) imply
$$
\Vert\Phi^{i}_x(\cdot,t)\Vert^2_0=\frac1{\pi}
\int\limits_{\mathbb{T}}
\sin^2(x\theta)\left(|\hat{\mathcal{G}}^{i0}_{t,+}(\theta)|^2
+|\hat{\mathcal{G}}^{i1}_{t,+}(\theta)|^2\right)\,d\theta,
\quad x\in\mathbb{Z}_+,\quad i=0,1.
$$
Hence, by (\ref{hatcalG}), we have
$\Vert\Phi^{1}_x(\cdot,t)\Vert^2_0\le C<\infty$ and
\begin{equation}
\label{3.12}
\Vert\Phi^{0}_x(\cdot,t)\Vert^2_0\le \int\limits_{\mathbb{T}}
\sin^2(x\theta)\left(C_1+C_2\frac{1}{\phi_+^2(\theta)}\right)\,d\theta
\le C_3+C_4x,\qquad x\in\mathbb{N},
\end{equation}
where the constants  $C_3$ and $C_4$ do not depend on $t\in\mathbb{R}$ and $x\in\mathbb{N}$.
Moreover, $C_4=0$ if $\kappa_+\not=0$, by (\ref{phi}).
If $\kappa_+=0$, then $\phi_+^2(\theta)=4\nu_+^2\sin^2(\theta/2)$
and the bound in the r.h.s. of (\ref{3.12}) follows from Fei\'er's theorem (see, e.g., \cite{Kat}).
$\Box$
\begin{cor}
Let
$\alpha<-1/2$ if $\kappa_-\kappa_+\not=0$, and $\alpha<-1$ otherwise.
Then
\begin{equation}
 \label{ubound}
\sup\limits_{t\in\mathbb{R}}
\int\Vert Y\Vert^2_{\alpha}\,\nu_t(dY)\le C<\infty.
\end{equation}
\end{cor}
Indeed, applying the bound~(\ref{Qnu}) gives
\begin{align*}
 \int\Vert Y\Vert^2_{\alpha}\,\nu_t(dY)=
 \sum\limits_{x\in\mathbb{Z}}\langle x\rangle^{2\alpha}
\left( Q_t^{\nu,00}(x,x)+Q_t^{\nu,11}(x,x)\right)
 \le \sum\limits_{x\in\mathbb{Z}}\langle x\rangle^{2\alpha}
\left(C_1+C_2|x|\right)\le C(\alpha)<\infty,
\end{align*}
by the choice of the $\alpha$.
\medskip

Introduce the limiting matrix $Q^{\nu}_\infty(x,y)$ by the rule
\begin{equation}\label{2.20}
Q^{\nu}_{\infty}(x,y)=\left\{
\begin{array}{lll}
Q_{\infty,+}(x,y)&\mbox{if }\,\,\, x,y>0,\\
 Q_{\infty,-}(x,y)&\mbox{if }\,\,\, x,y<0,\\
 0&\mbox{otherwise},
\end{array}\right.
\end{equation}
where
\begin{equation}\label{correlation}
Q_{\infty,\pm}(x,y):=q_{\infty,\pm}(x-y)-q_{\infty,\pm}(x+y)-q_{\infty,\pm}(-x-y)+
q_{\infty,\pm}(-x+y),\quad x,y\in\mathbb{Z}_\pm.
\end{equation}
The Fourier transforms of the entries of  $q_{\infty,\pm}(x)$, $x\in\mathbb{Z}$,
 have the form
\begin{equation}\label{qinfty}
\begin{array}{lll}
&\hat q_{\infty,\pm}^{00}(\theta)=
\frac14\left(\hat q_\pm^{00}(\theta)+\hat q_\pm^{11}(\theta)\phi_\pm^{-2}(\theta)\right)
\pm \frac{i}{4}\,{\rm sign}(\theta)\phi_\pm^{-1}(\theta)\left(\hat q_\pm^{10}(\theta)-\hat q_\pm^{01}(\theta)\right),\\
&  \hat q^{11}_{\infty,\pm}(\theta)=\phi_\pm^2(\theta)\hat q_{\infty,\pm}^{00}(\theta),\\
&\hat q_{\infty,\pm}^{01}(\theta)=-\hat q_{\infty,\pm}^{10}(\theta)=
\pm i\,{\rm sign}(\theta)\phi_\pm(\theta)\hat q_{\infty,\pm}^{00}(\theta),
\end{array}\end{equation}
where $\theta\in\mathbb{T}$ if $\kappa_\pm\not=0$ and $\theta\in\mathbb{T}\setminus\{0\}$ otherwise,
the functions $q_\pm^{ij}$, $i,j=0,1$,  are the entries of the matrices $q_\pm$
from  condition~(\ref{1.7'}),
 $\phi_\pm(\theta)$ are defined in (\ref{phi}).
\medskip

{\bf Remark}. 
{\em By (\ref{2.7}), $\hat q^{ii}_{\pm}(-\theta)=\hat q^{ii}_{\pm}(\theta)$
 and $\hat q^{01}_{\pm}(-\theta)=\hat q^{10}_{\pm}(\theta)$.
 Then, (\ref{qinfty}) gives
 \begin{equation*}\label{3.17}
 \hat q^{ii}_{\infty,\pm}(-\theta)=\hat q^{ii}_{\infty,\pm}(\theta),
 \qquad
 \hat q^{ij}_{\infty,\pm}(-\theta)=
  \hat q^{ji}_{\infty,\pm}(\theta)=
  -\hat q^{ij}_{\infty,\pm}(\theta)\quad \mbox{if }\,\,i\not=j,
 \quad i,j=0,1.
  \end{equation*}
 Therefore, by (\ref{2.20}) and (\ref{correlation}),
$Q^{ij}_{\infty,\pm}(x,y)=0$ and
$Q^{\nu,ij}_{\infty}(x,y)=0$  if $i\not=j$,
}
\begin{equation*}
Q^{ii}_{\infty,\pm}(x,y)=\frac2{\pi}\int\limits_{\mathbb{T}}\hat q^{ii}_{\infty,\pm}(\theta)
\sin(x\theta)\sin(y\theta)\,d\theta=Q^{\nu,ii}_{\infty}(x,y),\qquad x,y\in\mathbb{Z}_\pm.
\end{equation*}

Denote by ${\cal Q}^\nu_t (\Psi,\Psi)$, $t\in\mathbb{R}$,  a  real-valued quadratic form
on ${\cal S}=[S(\mathbb{Z})]^2$ with the matrix kernel $Q^\nu_t(x,y)$.
Using (\ref{2.20}), we have
\begin{equation}\label{qpp}
{\cal Q}^\nu_{\infty} (\Psi,\Psi)=\langle Q^\nu_{\infty}(x,y),\Psi(x)\otimes \Psi(y)\rangle
=\sum_\pm \langle Q_{\infty,\pm}(x,y),\Psi(x)\otimes \Psi(y)\rangle_\pm.
 \end{equation}

The following theorem was proved in \cite{D20}.
\begin{theorem}\label{l1}
Let $\alpha<-1/2$ if $\kappa_-\kappa_+\not=0$, and $\alpha<-1$ otherwise.
 Then the following assertions hold.
(i) Let conditions {\bf S1}--{\bf S3} be fulfilled. Then
 the correlation functions of the measures $\nu_t$ converge to a limit:
\begin{equation*}
 Q^{\nu}_{t}(x,y):=\int\Big( Y(x)\otimes Y(y)\Big)\,\nu_{t}(dY)\to
 Q^{\nu}_{\infty}(x,y), \qquad t\to\infty,\qquad x,y\in\mathbb{Z},
\end{equation*}
where the limiting correlation matrix $ Q^{\nu}_{\infty}(x,y)$ is of the form~(\ref{2.20}).

(ii)
Let conditions {\bf S1}, {\bf S3} and {\bf S4}  be fulfilled.
Then the measures $\nu_{t}$ converge weakly to a limit measure as $t\to\infty$ on the space
 ${\cal H}_{\alpha}$.
The limit measure $\nu_{\infty}$ is Gaussian  with zero mean value
and its  characteristic functional is
 $$
 \hat \nu_\infty(\Psi):=
 \int e^{i\langle Y,\Psi\rangle}\nu_\infty(dY)
 =\exp\left\{ -\frac12 {\cal Q}^\nu_{\infty} (\Psi,\Psi)\right\},
 \quad \Psi\in{\cal S},
 $$
where the quadratic form ${\cal Q}^\nu_{\infty}$ is defined in (\ref{qpp}).
 \end{theorem}

Below we will use an additional property of the  quadratic form
${\cal Q}^\nu_t$, $t\in\mathbb{R}$. To state it we first introduce auxiliary spaces.
\begin{definition}
For any sequence $\psi$, we introduce odd sequences $\psi_-$ and $\psi_+$ by the rule
\begin{equation}\label{3.13}
\psi_\pm(x)=\left\{\begin{array}{ll}
\psi(x)&\mbox{for }~\pm x>0,\\
0&\mbox{for }~x=0,\\
-\psi(-x)&\mbox{for }~\pm x<0.\\
\end{array}\right.
\end{equation}
Define the Hilbert space
$\ell^2(\kappa):=\{\psi\in\ell^2:\hat \psi_+\phi_+^{-1}(\theta),\hat\psi_-\phi_-^{-1}(\theta)\in L^2(\mathbb{T})\}$
with the norm
$$
\Vert\psi\Vert_{\ell^2(\kappa)}:=\Vert\psi\Vert_{\ell^2}+
\sum\limits_{\pm}\left\Vert F^{-1}_{\theta\to x}[\phi_\pm^{-1}(\theta)]*\psi_\pm\right\Vert_{\ell^2}.
$$
Introduce the space $\mathcal{H}{(\kappa)}:=\ell^2(\kappa)\times\ell^2$
with the norm
$$
\Vert\Psi\Vert_{\mathcal{H}(\kappa)}:=\Vert\Psi^0\Vert_{\ell^2(\kappa)}+
\Vert\Psi^1\Vert_{\ell^2},\qquad \Psi=(\Psi^0,\Psi^1).
$$
In particular, if $\kappa_-\kappa_+\not=0$, then $\mathcal{H}{(\kappa)}=\mathcal{H}_0=\ell^2\times\ell^2$.
\end{definition}

{\bf Remark}.
{\em By (\ref{3.13}), in the Fourier transform,
$|\widehat{\psi}_\pm(\theta)|\le C|\sin\theta|\sum\limits_{\pm x>0}|x||\psi(x)|$, $\theta\in\mathbb{T}$.
Hence, if $\sum\limits_{x\in\mathbb{Z}}|x||\psi(x)|<\infty$, then
$ \widehat{\psi}_\pm\phi_\pm^{-1}\in C(\mathbb{T})$.
In particular, $\ell^2_{-\alpha}\subset \ell^2(\kappa)$ for $\alpha<-3/2$, since
$\sum\limits_{x\in\mathbb{Z}}|x||\psi(x)|\le C\Vert \psi\Vert_{-\alpha}$
by the Cauchy--Schwartz inequality.
}
\begin{lemma}\label{l3.7}
The quadratic forms $\mathcal{Q}_t^\nu(\Psi,\Psi)$ and
the characteristic functionals $\hat\nu_t(\Psi)$, $t\in\mathbb{R}$,
are equicontinuous in $\mathcal{H}(\kappa)$.
\end{lemma}
{\bf Proof}\,
For $t\in\mathbb{R}$, introduce a formal adjoint operator $U'_0(t)$ to the
solving operator $U_0(t)$,
$$
\langle U_0(t) Y,\Psi\rangle=\langle Y, U'_0(t) \Psi\rangle,\qquad
Y\in \mathcal{H}_\alpha,\quad \Psi\in \mathcal{S}.
$$
Then, the action of the group $U'_0(t)$ coincides with the action
of $U_0(t)$ up to the order of the components. Namely,
$U'_0(t) \Psi=\left(\dot\psi(\cdot,t),\psi(\cdot,t)\right)$,
where $\psi(x,t)$ is a solution to problem~(\ref{a.1})
with the initial data $(u_0,v_0)=(\Psi^1,\Psi^0)$.
Using (\ref{c4.1}), we have
$$
\mathcal{Q}_t^\nu(\Psi,\Psi)=
\mathcal{Q}_0^\nu\left(U'_0(t)\Psi,U'_0(t)\Psi\right)\le
C\Vert U'_0(t)\Psi\Vert_0^2.
$$
On the other hand, by (\ref{3.2'}) and (\ref{3.2}),
$$
\left(U'_0(t)\Psi\right)^j(y)=\sum\limits_{i=0}^1
\sum\limits_{x\in\mathbb{Z}_{\pm}} G^{ij}_{t,\pm}(x,y)\Psi^i(x)
=\sum\limits_{i=0}^1
\sum\limits_{x\in\mathbb{Z}} \mathcal{G}^{ij}_{t,\pm}(x-y)\Psi^i_\pm(x)\quad
\mbox{for }\,y\in\mathbb{Z}_\pm.
$$
Here we use notation~(\ref{3.13}).
Therefore, applying the Parseval identity and (\ref{hatcalG}), we obtain
\begin{equation*}
  \Vert U'_0(t)\Psi\Vert_0^2\le
  C\sum\limits_\pm\int\limits_{\mathbb{T}}\left((1+\phi_\pm^{-2}(\theta))
  |\hat\Psi^0_\pm(\theta)|^2+  |\hat\Psi^1_\pm(\theta)|^2\right)\,d\theta
  \le C\Vert\Psi\Vert^2_{\mathcal{H}(\kappa)}.
\end{equation*}
Hence,
\begin{equation}\label{3.15'}
\mathcal{Q}_t^\nu(\Psi,\Psi)\le C\Vert\Psi\Vert^2_{\mathcal{H}(\kappa)}
\qquad \mbox{uniformly in }\,\,t\in\mathbb{R}.
\end{equation}
This implies the equicontinuity of the characteristic functionals $\hat\nu_t(\Psi)$,
$t\in\mathbb{R}$.
Indeed, by the Cauchy--Schwartz inequality and (\ref{3.15'}), one obtains
\begin{align*}
\left|\hat \nu_t(\Psi_1)-\hat \nu_t(\Psi_2)\right|&
=\Big|\int\left(e^{i\langle Y,\Psi_1\rangle}-e^{i\langle Y,\Psi_2\rangle}\right)\,\nu_t(dY)\Big|
\le \int\left|e^{i\langle Y,\Psi_1-\Psi_2\rangle}-1\right|\,\nu_t(dY)\\
&\le \int\left|\langle Y,\Psi_1-\Psi_2\rangle\right|\,\nu_t(dY)
\le\sqrt{\int\left|\langle Y,\Psi_1-\Psi_2\rangle\right|^2\,\nu_t(dY)}\\
&=\sqrt{\mathcal{Q}_t^\nu(\Psi_1-\Psi_2,\Psi_1-\Psi_2)}\le C\Vert \Psi_1-\Psi_2\Vert_{\mathcal{H}(\kappa)}.
\quad\Box
\end{align*}
\begin{definition}
(i)
 Denote by $R_0$ the operator $R_0:{\cal H}_{\alpha}\to\mathfrak{C}^1_{\alpha}$ such that
\begin{equation*}
(R_0Y_0)(x,t)=z(x,t),
\end{equation*}
where $z(x,t)$ is the solution to problem~(\ref{a.1}) with the initial data $Y_0=
(Y_0^0,Y_0^1)\equiv(u_0,v_0)$.
Then, by (\ref{3.2'}),
\begin{equation}\label{R_0}
(R_0Y_0)(x,t)=\left\{
\begin{array}{lll}
\sum\limits_{j=0,1}\sum\limits_{y\in\mathbb{Z}_+}
 G^{0j}_{t,+}(x,y) Y_{0}^j(y) &\mbox{for }& x\in\mathbb{Z}_+,\\
 \sum\limits_{j=0,1}\sum\limits_{y\in\mathbb{Z}_-}
 G^{0j}_{t,-}(x,y) Y_{0}^j(y)& \mbox{for }& x\in\mathbb{Z}_-.
\end{array}
\right.
\end{equation}
In particular,
 $\left(R_0Y_0\right)(0,t)=0$ for all $t$.

(ii) Introduce a Borel probability measure $P^\nu$
on the space $\mathfrak{C}^1_{\alpha}$ as
$$
P^\nu(\omega)=\nu_0(R_0^{-1}\omega),  \quad \forall\omega\in{\cal B}(\mathfrak{C}^1_\alpha).
$$
This measure is called a space-time statistical solution to problem (\ref{a.1}).

(iii) Denote by $\{P^\nu_\tau, \tau\in\mathbb{R}\}$ the family of measures defined by the rule
$$
P^\nu_\tau(\omega)=P^\nu(S_\tau^{-1}\omega),
\quad
 \forall\omega\in{\cal B}({\mathfrak C}^1_\alpha),\quad \tau\in\mathbb{R}.
 $$
\end{definition}

In this section, we prove the following theorem.
 \begin{theorem}\label{l2.1}
 Let
$\alpha<-1/2$ if $\kappa_-\kappa_+\not=0$, and $\alpha<-1$ otherwise.
 Then the following assertions hold.
(i) Let conditions {\bf S1} and {\bf S2} be fulfilled.
 Then the bounds are true:
\begin{equation}\label{4.1}
\sup_{\tau\ge0} \int \br z\br^2_{\alpha,1,T} P^\nu_\tau(dz) \le C(\alpha)<\infty,\qquad\forall\, T>0,
\end{equation}
where the constant $C(\alpha)$ does not depend on $T>0$.

(ii) Let conditions {\bf S1}--{\bf S3} be fulfilled.
Then for any $v_1,v_2\in\mathcal{P}$,
\begin{equation}\label{3.16}
\mathcal{Q}_{\tau}^{P,\nu}(v_1,v_2):=\int[z,v_1][z,v_2]\,P^\nu_\tau(dz)
\to \mathcal{Q}_\infty^{P,\nu}(v_1,v_2),
\qquad \tau\to\infty.
\end{equation}
Here
\begin{equation}\label{3.14}
\mathcal{Q}_\infty^{P,\nu}(v_1,v_2):=[Q_\infty^{P,\nu},v_1\otimes v_2],
 \end{equation}
where the limiting correlation matrix $Q_\infty^{P,\nu}$ is of a form
\begin{equation}\label{100}
Q^{P,\nu}_{\infty}(x_1,x_2,t_1,t_2)=\left\{
\begin{array}{lll}
Q^{P,\nu}_{\infty,+}(x_1,x_2,t_1,t_2)&\mbox{if }\,\,\, x_1,x_2>0,\\
 Q^{P,\nu}_{\infty,-}(x_1,x_2,t_1,t_2)&\mbox{if }\,\,\, x_1,x_2<0,\\
 0&\mbox{otherwise},
\end{array}\right. \quad t_1,t_2\in \mathbb{R}.
 \end{equation}
Here
\begin{equation}\label{101}
Q^{P,\nu}_{\infty,\pm}(x_1,x_2,t_1,t_2):=
\frac{2}{\pi}\int\limits_{\mathbb{T}}
\cos\left(\phi_\pm(\theta)(t_1-t_2)\right)
\hat q^{00}_{\infty,\pm}(\theta)
\sin(x_1\theta)\sin(x_2\theta)\,d\theta,
\end{equation}
where $\hat q^{00}_{\infty,\pm}$ is  defined in (\ref{qinfty}).

(iii) Let conditions {\bf S1}, {\bf S3} and {\bf S4} be fulfilled.
Then the measures $P_\tau^\nu$ converge weakly to a limiting measure
$P_\infty^\nu$ on the space $\mathfrak{C}^0_\alpha$ as $\tau\to\infty$.
The characteristic functional of $P_\infty^\nu$ is
\begin{equation}\label{3.21}
\hat {P}^\nu_\infty (v)\equiv
\int e^{i[z,v]}P^\nu_\tau(dz)=
\exp\left\{-\frac{1}{2}  \mathcal{Q}^{P,\nu}_{\infty}(v,v)\right\},\qquad
v \in \mathcal{P},
\end{equation}
where the quadratic form $\mathcal{Q}^{P,\nu}_{\infty}$ is defined in (\ref{3.14})--(\ref{101}).
 \end{theorem}
{\bf Proof}\,
(i) At first, note that
\begin{equation}\label{2.23-0}
P^\nu_\tau(\omega)=\nu_\tau(R_0^{-1}\omega)\quad
\mbox{ for any }\,\,\omega\in {\cal B}(\mathfrak{C}^1_\alpha)\quad \mbox{and }\,\,\tau>0,
\end{equation}
where $\nu_\tau$ is defined in Definition~\ref{mu0}.
Hence, the bound~(\ref{4.1}) follows from (\ref{ubound}), because
\begin{align*}
\int \br z\br^2_{\alpha,1,T} P^\nu_\tau(dz)&=
\int \br R_0Y\br^2_{\alpha,1,T} \,\nu_\tau(dY)=
\sup_{|s|\le T}\int \Vert U_0(s)Y\Vert^2_{\alpha}\, \nu_\tau(dY)\\
&
=\sup_{|s|\le T}\int \Vert Y\Vert^2_{\alpha}\, \nu_{s+\tau}(dY)
\le C(\alpha)<\infty.
\end{align*}

(ii) Let $z\equiv z(\cdot,t)$ be a solution to problem~(\ref{a.1}). Then, for any $v\in \mathcal{P}$,
\begin{equation}\label{4.2}
[z,v]=[R_0 Y_0, v]=\langle Y_0, R'_0 v\rangle,
\end{equation}
where $R'_0$ is an adjoint operator to the operator $R_0$,
$R'_0 v=\left((R'_0 v)^0,(R'_0 v)^1\right)$, and
\begin{equation}\label{102}
(R'_0 v)^j(y)=
\left\{
\begin{array}{ll}
\sum\limits_{x\in\mathbb{Z}_+}\int\limits_{-\infty}^{+\infty} G^{0j}_{t,+}(x,y)v(x,t)\,dt&\mbox{if }\,\,y\in\mathbb{Z}_+,\\
\sum\limits_{x\in\mathbb{Z}_-}\int\limits_{-\infty}^{+\infty} G^{0j}_{t,-}(x,y)v(x,t)\,dt&\mbox{if }\,\,y\in\mathbb{Z}_-,
\end{array}\right.\quad j=0,1,
\end{equation}
by (\ref{R_0}). In particular, using (\ref{3.2}),
 we have $(R'_0 v)^j(0)=0$.
Below we use the notation
$$
\Vert v\Vert_{L^1(\mathbb{R};X)}:=\int\limits_{-\infty}^{+\infty}
\Vert  v(\cdot,t)\Vert_{X}\,dt
\qquad \mbox{for }\,\, v(\cdot,t)\in L^1(\mathbb{R};X)\quad
\mbox{with }\,\, X=\ell^2\quad \mbox{or }\,\,X=\ell^2(\kappa).
$$
We state the additional properties of the operator $R'_0$ in the following lemma.
\begin{lemma}\label{l3.11}
(i) If $\kappa_-\kappa_+\not=0$, then $R'_0 v\in\mathcal{S}$ for any $v\in\mathcal{P}$.
(ii) For any $v\in L^1(\mathbb{R}; \ell^2(\kappa))$,
\begin{equation}\label{3.26'}
\Vert R'_0 v\Vert_{\mathcal{H}(\kappa)}\le C\Vert v\Vert_{L^1(\mathbb{R};\ell^2(\kappa))}.
\end{equation}
\end{lemma}
{\bf Proof}\, The first assertion follows from (\ref{102}) and formulas (\ref{3.2})--(\ref{phi}).
To prove assertion~(ii), we apply (\ref{102}), notation~(\ref{3.13}) for $v(x,t)$, and equations~(\ref{3.2}) and obtain
$$
(R'_0 v)^j(y)=\sum\limits_{x\in\mathbb{Z}}\,\,\int\limits_{-\infty}^{+\infty}
\mathcal{G}^{0j}_{t,\pm}(x-y)v_\pm(x,t)\,dt\quad \mbox{for }\,\,y\in\mathbb{Z}_\pm.
$$
Hence, by the Parseval identity and (\ref{hatcalG}), we have
$$
\Vert(R'_0 v)^j\Vert_{\ell^2}\le\sum\limits_{\pm}
\int\limits_{-\infty}^{+\infty}
\Vert \hat{\mathcal{G}}^{0j}_{t,\pm}(\theta)\hat v_\pm(\theta,t)\Vert_{L^2(\mathbb{T})}\,dt\le
\sum\limits_{\pm}
\int\limits_{-\infty}^{+\infty} \Vert \phi_\pm^{-j}(\theta)\hat v_\pm(\theta,t)\Vert_{L^2(\mathbb{T})}\,dt.
$$
Therefore,
$\Vert(R'_0 v)^0\Vert_{\ell^2}\le C\Vert v\Vert_{L^1(\mathbb{R};\ell^2)}$.
If $\kappa_-\kappa_+\not=0$, then
the same bound is valid for $(R'_0 v)^1$.
If $\kappa_-\kappa_+=0$, then
$\Vert(R'_0 v)^1\Vert_{\ell^2}\le C
\Vert v\Vert_{L^1(\mathbb{R};\ell^2(\kappa))}$.
Furthermore,
\begin{align*}
\Vert(R'_0 v)^0\Vert_{\ell^2(\kappa)}&=
\Vert(R'_0 v)^0\Vert_{\ell^2}+\sum\limits_{\pm}
\left\Vert F^{-1}[\phi_\pm^{-1}] * \left((R'_0 v)^0\right)_\pm
\right\Vert_{\ell^2}\\
&\le \int\limits_{-\infty}^{+\infty}
\Vert v(\cdot,t)\Vert_{\ell^2}\,dt+
\sum\limits_{\pm}
\int\limits_{-\infty}^{+\infty} \Vert \phi_\pm^{-1}(\theta)\hat v_\pm(\theta,t)\Vert_{L^2(\mathbb{T})}\,dt.
\end{align*}
This implies the bound (\ref{3.26'}).
In particular, $R'_0 v\in \mathcal{H}(\kappa)$
for any $v\in\mathcal{P}$.
Lemma~\ref{l3.11} is proved.  $\Box$
\medskip

We return to the proof of assertion~(ii) of Theorem~\ref{l2.1}.
For any $v_1,v_2\in\mathcal{P}$,
\begin{equation*}
\begin{split}
\mathcal{Q}_{\tau}^{P,\nu}(v_1,v_2)&=
\int[R_0Y,v_1][R_0Y,v_2]\,\nu_\tau(dY)=\int\langle Y,R'_0 v_1\rangle\langle Y,R'_0 v_2\rangle\,\nu_\tau(dY)\\
&=\langle Q^\nu_\tau(x,y),(R'_0 v_1)(x)\otimes (R'_0 v_2)(y)\rangle
\equiv\mathcal{Q}_{\tau}^{\nu}(R'_0 v_1,R'_0 v_2).
\end{split}
\end{equation*}
If $\kappa_-\kappa_+\not=0$, then  $R'_0v_i\in{\cal S}$  and
Theorem~\ref{l1}~(i) implies
\begin{equation}\label{3.27'}
\mathcal{Q}_{\tau}^{P,\nu}(v_1,v_2)=
\mathcal{Q}_{\tau}^{\nu}(R'_0 v_1,R'_0 v_2)\to
\mathcal{Q}_{\infty}^{\nu}(R'_0 v_1,R'_0 v_2),\qquad \tau\to\infty.
\end{equation}
If $\kappa_\pm=0$, then convergence~(\ref{3.27'})
follows from Theorem~\ref{l1}~(i) and Lemmas~\ref{l3.7} and \ref{l3.11},
because the space $\mathcal{S}$ is dense in $\mathcal{H}(\kappa)$.
\smallskip

It remains to check formula~(\ref{101}).
Using (\ref{3.27'}), (\ref{qpp}), (\ref{2.20}) and (\ref{102}), we have
\begin{align}
\mathcal{Q}_{\infty}^{P,\nu}(v_1,v_2)&=
\mathcal{Q}_{\infty}^{\nu}(R'_0v_1,R'_0v_2)
=\langle Q^\nu_\infty(y_1,y_2),R'_0v_1(y_1)\otimes R'_0v_2(y_2)\rangle\notag\\
&=\sum\limits_\pm\sum\limits_{x_1,x_2\in\mathbb{Z}_{\pm}}
\int\limits_{-\infty}^{+\infty}dt_1\int\limits_{-\infty}^{+\infty}
Q^{P,\nu}_{\infty,\pm}(x_1,x_2,t_1,t_2)v_1(x_1,t_1)v_2(x_2,t_2)\,dt_2, \notag
\end{align}
where, by definition,
$$
Q^{P,\nu}_{\infty,\pm}(x_1,x_2,t_1,t_2):=
\sum\limits_{i,j=0,1}\sum\limits_{y_1,y_2\in\mathbb{Z}_\pm}
Q^{ij}_{\infty,\pm}(y_1,y_2) G^{0i}_{t_1,\pm}(x_1,y_1)G^{0j}_{t_2,\pm}(x_2,y_2)
$$
for $x_1,x_2\in\mathbb{Z}_\pm$, $t_1,t_2\in\mathbb{R}$. Hence, (\ref{100}) holds.
Using formulas (\ref{3.2}) and (\ref{correlation}) and
the Parseval identity, we obtain
\begin{align}
Q^{P,\nu}_{\infty,\pm}(x_1,x_2,t_1,t_2)&=
\sum\limits_{i,j=0,1}\sum\limits_{y_1,y_2\in\mathbb{Z}}
q^{ij}_{\infty,\pm}(y_1-y_2) G^{0i}_{t_1,\pm}(x_1,y_1)G^{0j}_{t_2,\pm}(x_2,y_2)\notag\\
&=\frac{4}{2\pi}\sum\limits_{i,j=0,1}
\int\limits_{\mathbb{T}}\hat q^{ij}_{\infty,\pm}(\theta)
\hat{\mathcal{G}}^{0i}_{t_1,\pm}(\theta) \hat{\mathcal{G}}^{0j}_{t_2,\pm}(\theta)
\sin(x_1\theta)\sin(x_2\theta)\,d\theta \label{3.20}
\end{align}
for $\pm x_1,\pm x_2>0$, $t_1,t_2\in\mathbb{R}$.
Applying (\ref{qinfty}) and (\ref{hatcalG}),
we obtain
\begin{align*}
Q^{P,\nu}_{\infty,\pm}(x_1,x_2,t_1,t_2)=&
\frac{2}{\pi}\int\limits_{\mathbb{T}}
\left\{\cos\left(\phi_\pm(\theta)(t_1-t_2)\right)
\hat q^{00}_{\infty,\pm}(\theta)
-\frac{\sin\left(\phi_\pm(\theta)(t_1-t_2)\right)}{\phi_\pm(\theta)}
\hat q^{01}_{\infty,\pm}(\theta)\right\}\\
&\times\sin(x_1\theta)\sin(x_2\theta)\,d\theta.
\end{align*}
This implies relation~(\ref{101})
since the functions $\hat q^{01}_{\infty,\pm}(\theta)$ are odd and $\phi_\pm(\theta)$ are even.
\medskip

(iii) According to the methods of \cite{VF}, to establish
the weak convergence of the measures $P_\tau^\nu$ on the space $\mathfrak{C}^0_\alpha$
it is enough to prove the following two assertions:
\begin{itemize}
\item[(A1)] {\em The family of measures $\{P_\tau^\nu,\tau\in\mathbb{R}\}$
is weakly compact in  $\mathfrak{C}^0_\alpha$};
\item[(A2)] {\em The characteristic functionals of $P_\tau^\nu$ converge to a limit as $\tau\to\infty$}.
\end{itemize}
The first (second) assertion provides the existence (resp., uniqueness)
of the limit measures  $P^\nu_\infty$.
\medskip

{\bf Proof of assertion~(A1)}:
To prove the weak compactness of the family $\{P^\nu_\tau,\,\tau\in \mathbb{R}\}$,
we verify that this family satisfies
the following conditions~(a) and (b)
of the Prokhorov Theorem (see, e.g., \cite{GS}):
\begin{itemize}
\item[(a)] $\sup \{P^\nu_\tau,\,\tau\in \mathbb{R}\} <\infty$,

\item[(b)] for any $\varepsilon>0$
there is a compact $K_\varepsilon$ in $\mathfrak{C}^0_\beta$ such that
$\sup_\tau P^\nu_\tau(\mathfrak{C}^0_\beta\setminus K_\varepsilon)<\varepsilon$.
\end{itemize}
Condition~(a)  holds since $P^\nu_\tau$ are probability measures.
To check  condition~(b), we apply the technique of \cite[Theorem~XII.5.2]{VF}.  
For $k=0,1$ and $T>0$, denote by $\mathfrak{C}^k_{\alpha,T}$
the space of the functions $t\to u(\cdot,t)\in\ell^2_\alpha$, $t\in[0,T]$,
for which the norm (\ref{1.3}) is finite.
For any $T>0$ and $M>0$, introduce sets
$$
K(T,M):=\{u\in\mathfrak{C}^1_{\alpha,T}:\br u\br_{\alpha,1,T}\le M\}.
$$
Below we choose $M\equiv M(T)$ by a special way.
The sets $K(T,M)$ are uniformly bounded and uniformly equicontinuous. Since
the embedding of the spaces $\ell^2_{\alpha}$ in $\ell^2_{\beta}$ is compact if $\alpha>\beta$, then
the sets $K(T,M)$ are precompact in $\mathfrak{C}^0_{\beta,T}$ by the
Dubinskii embedding theorems~(see, e.g., \cite{Dub} or \cite[Theorem~IV.4.1]{VF}) 
using  the Arzel\`a--Ascoli theorem (see, e.g., \cite[Ch.3, \S~3]{Yo}).
For $T>0$, introduce the operator $J_T:\mathfrak{C}^0_{\beta}\to\mathfrak{C}^0_{\beta,T}$
of the restriction of the functions $u(x,t)\in\mathfrak{C}^0_{\beta}$
from $\mathbb{Z}\times \mathbb{R}$ into $\mathbb{Z}\times [-T,T]$.
Applying the Chebyshev inequality and the bound~(\ref{4.1}), we obtain
\begin{equation}\label{4.6}
P^\nu_\tau\{\mathfrak{C}^0_\beta\setminus J_T^{-1} \overline{K(T,M)}\}
\le \int\br u\br^2_{\alpha,1,T}\, P^\nu_\tau(du)/M^2\le C(\alpha)/M^2,
\end{equation}
where by $\overline{K}$ we denote the closure of $K$
in the topology of the metrizable space $\mathfrak{C}^0_\beta$.
For any $\varepsilon>0$, we choose the positive constants $M=M_\varepsilon(T)$ such that
$$
C(\alpha)\sum\limits_{T=1}^\infty\frac{1}{M^2_\varepsilon(T)}<\varepsilon.
$$
Set $K_\varepsilon:=\bigcap\limits_{T=1}^\infty J_T^{-1}\overline{K(T,M_\varepsilon(T))}$.
Then the bound~(\ref{4.6}) implies the condition~(b).
\medskip

{\bf Proof of assertion~(A2)}:
  Applying (\ref{2.23-0}), (\ref{4.2}), Theorem~\ref{l1}~(ii)
  and Lemmas~\ref{l3.7} and \ref{l3.11},
  we obtain that   for every $v\in \mathcal{P}$,
\begin{equation*}
\hat {P}^\nu_\tau (v)=
\int e^{i[z,v]}P^\nu_\tau(dz)=
\int e^{i\langle Y_0,R'_0v\rangle} \nu_\tau(dY_0)
\to\exp\left\{-\frac12 {\cal Q}_\infty^\nu(R'_0v,R'_0v)\right\},\quad \tau\to\infty.
\end{equation*}
The assertion~(iii) of Theorem~\ref{l2.1} is proved.$\Box$


\subsection{Perturbed problem}\label{sec3.2}
The key role in the proof of convergence~(\ref{1.8i}) for problem~(\ref{1}) plays
the following lemma.
\begin{lemma}\label{l3.10}   (see \cite[Lemma~4.3]{D20})
Let $Y_0\in{\cal H}_{\alpha}$, $\alpha<-3/2$, and conditions~{\bf C},
{\bf S1}, and {\bf S2} hold.
Then there exists a linear bounded operator
$\Omega:\mathcal{H}_{0}\to \mathcal{H}_{\alpha}$
such that the following representation holds
\begin{equation}\label{3.29}
U(t)Y_0(x)= \Omega (U_0(t) Y_0)(x) +\delta(x,t),
\qquad\mbox{where }\,\,\, \mathbb{E}\Vert \delta(\cdot,t)\Vert_{\alpha}^2\le C\langle t\rangle^{-1}.
\end{equation}
Here   $U(t)Y_0\equiv(u(\cdot,t),\dot u(\cdot,t))$
is a solution to problem~(\ref{1})--(\ref{4}),
the operator $\Omega$ is of the form
\begin{equation}\label{operatorOmega}
\Omega Y=Y+\Gamma Y,\qquad
(\Gamma Y)(x):=\left(\langle Y,\bar{\mathbf{\Gamma}}^0(x,\cdot)\rangle,
\langle Y,\bar{\mathbf{\Gamma}}^1(x,\cdot)\rangle\right),\qquad x\in\mathbb{Z},
\end{equation}
where $\bar{\mathbf{\Gamma}}^j(x,y)$, $j=0,1$, is a vector-valued function of the form
\begin{eqnarray*}
\bar{\mathbf{\Gamma}}^j(x,y)
=\left\{\begin{array}{lll}
\int\limits_0^{+\infty}\Gamma^\pm_x(s)\Big(U'_0(-s)\overline{\mathbf{G}}^j\Big)(y)\,ds&\mbox{if }\,\,\pm x>0,&\\
\overline{\mathbf{G}}^j(y)&\mbox{if }\,\,x=0,
\end{array}\right.\quad y\in\mathbb{Z}.
\end{eqnarray*}
Here
$\overline{\bf G}^j(y):=\nu^2_\pm {\bf G}^j_\pm(y)$ for $\pm y\ge0$,
\begin{equation*}
{\bf G}^j_\pm(y):=
\int\limits_0^{+\infty} N^{(j)}(s) {\bf g}_\pm^{0}(y,-s)\,ds,\quad y\in\mathbb{Z},
\qquad
{\bf g}^{0}_\pm(y,t):=\Big(
 G^{00}_{t,\pm}(\pm1,y), G^{01}_{t,\pm}(\pm1,y)\Big),
 \end{equation*}
 $N^{(0)}(s)\equiv N(s)$, $N^{(1)}(s)\equiv \dot N(s)$,
 the functions $N(s)$ and $\Gamma^\pm_x(s)$
are constructed in \cite{D20}.
They satisfy the following bounds:
$$
|N(s)|\le C\langle s\rangle^{-3/2},\qquad
\sum\limits_{x\in\mathbb{Z}_\pm\setminus\{0\}}\langle x\rangle^{2\alpha}
|\Gamma^\pm_x(s)|^2\le C\langle s\rangle^{-3},\quad s\in\mathbb{R},\quad \alpha<-3/2.
$$\end{lemma}
\begin{cor}\label{cor3.14}
 Let $\alpha<-3/2$. Then there is a bounded linear operator
 $\Omega':\mathcal{H}_{-\alpha}\to \mathcal{H}_0$ such that
for any $\Psi\in{\cal S}$ we have
\begin{equation} \label{6.21}
\langle U(t)Y_0,\Psi\rangle=\langle U_0(t)Y_0,\Omega'\Psi\rangle+\delta(t),
\qquad\mbox{where }\,\,\,
\mathbb{E}|\delta(t)|^2\le C\langle t\rangle^{-1}\Vert\Psi\Vert^2_{-\alpha}.
\end{equation}
The operator $\Omega'$ is  of a form
\begin{equation}\label{operatorOmega'}
\Omega'\Psi=\Psi +\Gamma'\Psi,\qquad
(\Gamma'\Psi)(y):=\sum\limits_{j=0}^1\langle\bar{\mathbf{\Gamma}}^j(\cdot,y),\Psi^j(\cdot)\rangle,
\qquad \Psi=(\Psi^0,\Psi^1).
\end{equation}
\end{cor}
{\bf Remark}.
{\em  As shown in \cite{D20},
$\Vert \bar{\mathbf{\Gamma}}^j(x,\cdot)\Vert_0\in \mathcal{H}_{\alpha}$   $\forall\alpha<-3/2$,
where $\Vert\cdot\Vert_0\equiv \Vert\cdot\Vert_{\mathcal{H}_0}$.
Furthermore, using the similar reasonings as in \cite{D20} one can check that
\begin{equation*}\label{3.34'}
 \Vert \bar{\mathbf{\Gamma}}^j(x,\cdot)\Vert_{\mathcal{H}(\kappa)}\in \mathcal{H}_{\alpha} \qquad
 \mbox{for any }\,\,\alpha<-3/2,\quad j=0,1.
\end{equation*}
Therefore,
\begin{equation}\label{3.36}
\Vert\Gamma'\Psi\Vert_{\mathcal{H}(\kappa)}\le C\Vert\Psi\Vert_{-\alpha}
\quad \mbox{and}\quad
\Vert\Omega'\Psi\Vert_{\mathcal{H}(\kappa)}\le C\Vert\Psi\Vert_{-\alpha}
\qquad \forall\Psi\in \mathcal{H}_{-\alpha}.
\end{equation}
} 

Before to prove Theorem~\ref{tA} we state the results concerning
the statistical solutions $\mu_t$ to problem~(\ref{1}).
\begin{definition}\label{def2.5}
$\mu_t$ is a Borel probability measure in
${\cal H}_\alpha$ which gives the distribution of $Y(t)$,
$\mu_t(B)=\mu_0(U(-t)B)$ for any  $B\in {\cal B}({\cal H}_\alpha)$, $t\in \mathbb{R}$.
The correlation functions of the  measure $\mu_t$ are  defined as
$$
Q_t^{ij}(x,y)= \mathbb{E} \left(Y^i(x,t)
 Y^j(y,t)\right),\quad i,j= 0,1,\quad \,x,y\in\mathbb{Z},\quad t\in\mathbb{R}.
$$
Here $Y^i(x,t)$ are the components of the  solution
$Y(t)=(Y^0(\cdot,t),Y^1(\cdot,t))=(u(\cdot,t),\dot u(\cdot,t))$.
Denote by   ${\cal Q}_t$  the quadratic form
with the matrix  kernel $(Q^{ij}_t(x,y))_{i,j=0,1}$,
$$
{\cal Q}_t (\Psi, {\Psi})=
\int\left|\langle Y,\Psi\rangle\right|^2\mu_t(dY)=
\sum\limits_{i,j=0,1}
\left\langle Q_t^{ij}(x,y),\Psi^i(x)\Psi^j(y)\right\rangle,\quad t\in\mathbb{R},\quad \Psi=(\Psi^0,\Psi^1)\in\mathcal{S}.
$$
\end{definition}
\begin{lemma}
Let $\alpha<-3/2$ and conditions~{\bf C}, {\bf S1}, and {\bf S2} be fulfilled.
Then the following bound holds
 \begin{equation}\label{2.18}
\sup_{t\in\mathbb{R}} \int \Vert Y\Vert^2_{\alpha}\,\mu_t(dY)=
\sup_{t\in\mathbb{R}} \mathbb{E} \Vert U(t)Y\Vert^2_{\alpha}
\le C<\infty.
\end{equation}
\end{lemma}
{\bf Proof}.
 As shown in \cite{D18}, for any $\alpha<-3/2$,
$\Vert U'_0(t)\bar{\mathbf{\Gamma}}^j(x,\cdot)\Vert_0\in \mathcal{H}_{\alpha}$
uniformly in $t\in\mathbb{R}$, i.e.,
\begin{equation}\label{3.33}
\sup\limits_{t\in\mathbb{R}}\left\Vert \left(\Vert U'_0(t)\bar{\mathbf{\Gamma}}^j(x,\cdot)\Vert_0\right)\right\Vert^2_{\alpha}
\equiv
\sup\limits_{t\in\mathbb{R}}\sum\limits_{x\in\mathbb{Z}}
 \langle x \rangle^{2\alpha}
 \left\Vert U'_0(t)\bar{\mathbf{\Gamma}}^j(x,\cdot)\right\Vert_0^2
<\infty.
\end{equation}
We check that
\begin{equation}\label{3.35}
\sup\limits_{t\in\mathbb{R}}\mathbb{E}\Vert\Omega U_0(t) Y_0\Vert^2_{\alpha}\le C<\infty.
\end{equation}
Indeed, applying (\ref{operatorOmega}), (\ref{ubound}),
 (\ref{c4.1}) and (\ref{3.33}) gives
\begin{align*}
\mathbb{E}\Vert\Omega U_0(t) Y_0\Vert^2_{\alpha}&\le
\mathbb{E}\Vert U_0(t)Y\Vert^2_{\alpha}+
\sum\limits_{j=0}^1\mathbb{E}\Vert
|\langle Y(\cdot), U'_0(t)\bar{\mathbf{\Gamma}}^j(x,\cdot)\rangle\Vert_{\alpha}^2\\
&\le C_1+
 \sum\limits_{j=0}^1\sum\limits_{x\in\mathbb{Z}}
 \langle x \rangle^{2\alpha}
 \mathcal{Q}_0\left(U'_0(t)\bar{\mathbf{\Gamma}}^j(x,\cdot),
 U'_0(t)\bar{\mathbf{\Gamma}}^j(x,\cdot)\right)\\
&\le C_1+
C_2 \sum\limits_{j=0}^1\sum\limits_{x\in\mathbb{Z}}
 \langle x \rangle^{2\alpha}
\Vert U'_0(t)\bar{\mathbf{\Gamma}}^j(x,\cdot)\Vert^2_0
\le C<\infty.
\end{align*}
Therefore, (\ref{3.29}) and (\ref{3.35}) imply the bound~(\ref{2.18}).
$\Box$
\begin{theorem}\label{t2.11} (see \cite[Theorems~2.3, 2.4]{D20})
 Let $\alpha<-3/2$ and condition {\bf C} hold.
 Then the following assertions are fulfilled.

  (i)    Let  conditions  {\bf S1}--{\bf S3} hold.
 Then  for all $\Psi\in{\cal S}$,
\begin{equation}\label{3.38}
\lim_{t\to\infty}\mathbb{E}|\langle Y(t),\Psi\rangle|^2
={\cal Q}_\infty(\Psi,\Psi)={\cal Q}^\nu_{\infty} (\Omega'\Psi,\Omega'\Psi),
\end{equation}
where the quadratic form ${\cal Q}^\nu_\infty$ is introduced in (\ref{qpp}).

 (ii)  Let conditions {\bf S1}, {\bf S3} and {\bf S4} hold.
Then the measures $\mu_t$ weakly converge to a Gaussian measure $\mu_\infty$ as $t\to\infty$
 on ${\cal H}_\alpha$.
 The characteristic functional of $\mu_\infty$ is of a form
$$
\hat\mu_\infty(\Psi)=\exp\{-{\cal Q}_\infty(\Psi,\Psi)/2\},
\quad \Psi\in{\cal S}.
$$
\end{theorem}
{\bf Remark}.
{\em It follows from the bounds~(\ref{3.15'}) and (\ref{3.36}) that
$$
\sup_{t\in\mathbb{R}}{\cal Q}^\nu_{t} (\Omega'\Psi,\Omega'\Psi)\le
C\Vert\Omega'\Psi\Vert^2_{\mathcal{H}(\kappa)}\le C_1\Vert\Psi\Vert^2_{-\alpha}
\qquad \forall\Psi\in \mathcal{H}_{-\alpha}.
$$
In particular, the r.h.s. of (\ref{3.38}) is defined  for any $\Psi\in \mathcal{S}$.
}
\medskip

 {\bf Proof of Theorem~\ref{tA}:}
At first, using Lemma~\ref{l3.10}, we estimate
$[S_\tau u,v]$, where $v\in\mathcal{P}$, $u\equiv u(x,t)$ is a solution
to problem~(\ref{1}), $S_\tau$ is defined in (\ref{1.5}).
Set $\vec{v}:=(v,0)$. For any $v\in\mathcal{P}$, we have
\begin{equation}\label{3.25}
 [S_\tau u,v]=[S_\tau z, T\Omega'\vec{v}]+\delta_\tau,
 \quad \mbox{where }\,\,
\mathbb{E}(\delta_\tau^2)=o(1),\quad  \tau\to\infty,
\end{equation}
 $z\equiv z(x,t)$ is a solution to problem~(\ref{a.1}),
and the operator $T$ is defined by the rule
\begin{equation}\label{operatorT}
T\Phi:=\Phi^0-\dot{\Phi}^1 \qquad \mbox{for }\,\,
\Phi\equiv \Phi(t)=(\Phi^0(t),\Phi^1(t)). 
\end{equation}
To prove (\ref{3.25}), we first write $[S_\tau u,v]$ in a form
\begin{equation}\label{3.26}
 [S_\tau u,v]=\int\limits_{-\infty}^{+\infty}
\langle U(t+\tau)Y_0,\vec{v}(\cdot,t)\rangle\, dt=
\int\limits_{-\infty}^{+\infty}
\langle U_0(t+\tau)Y_0,\Omega'\vec{v}(\cdot,t)\rangle\, dt+\delta_\tau,
\end{equation}
where $\mathbb{E}(\delta_\tau^2)=o(1)$, $\tau\to\infty$.
The bound~(\ref{3.26}) follows from Corollary~\ref{cor3.14}
 because
\begin{align}\label{3.45'}
\mathbb{E}(\delta_\tau^2)&\le \Big(\int\limits_{-\infty}^{+\infty}
\sqrt{\mathbb{E}\left|\langle U(t+\tau)Y_0,\vec{v}(\cdot,t)\rangle-
\langle U_0(t+\tau)Y_0,\Omega'\vec{v}(\cdot,t)\rangle\right|^2}
\,dt\Big)^2\notag\\
&\le C\Big(\int\limits_{-\infty}^{+\infty}
\langle t+\tau\rangle^{-1/2}\Vert v(\cdot,t)\Vert_{-\alpha}\,
dt\Big)^2=o(1)\quad \mbox{as }\,\,\tau\to\infty.
\end{align}
Secondly, we rewrite the integral in the r.h.s. of (\ref{3.26}) using
notation~(\ref{operatorT}):
$$
\int\limits_{-\infty}^{+\infty}\left\langle U_0(t+\tau)Y_0,\Omega'\vec{v}(\cdot,t)\right\rangle\,dt=
\int\limits_{-\infty}^{+\infty}\left\langle z(\cdot,t+\tau),T\Omega'\vec{v}(\cdot,t)\right\rangle\,dt
=[S_\tau z, T\Omega'\vec{v}].
$$
This implies representation~(\ref{3.25}).
Further, using (\ref{3.25}), we obtain that for $v_1,v_2\in\mathcal{P}$,
\begin{equation}\label{3.27}
\mathcal{Q}_\tau^P(v_1,v_2)=\int[u,v_1][u,v_2]\,P_\tau(du)
= \mathcal{Q}^{P,\nu}_\tau (T\Omega'\vec{v}_1,T\Omega'\vec{v}_2)+\delta'_\tau,
\end{equation}
where the quadratic form $\mathcal{Q}^{P,\nu}_\tau $
is introduced in (\ref{3.16}),
$\delta'_\tau=o(1)$ as $\tau\to\infty$.
Note that  $T\Omega'\vec{v}_i \not\in \mathcal{P}$, in general,
and we can not  apply convergence~(\ref{3.16}) immediately.

 At first, using the equality
$\mathcal{Q}^{P,\nu}_\tau (w_1,w_2)=\mathcal{Q}^{\nu}_\tau (R'_0w_1,R'_0w_2)$, we  obtain
$$
\mathcal{Q}_\tau^P(v_1,v_2)= \mathcal{Q}^{\nu}_\tau \left(R'_0T\Omega'\vec{v}_1,R'_0T\Omega'\vec{v}_2\right)+o(1),
\quad \tau\to\infty.
$$
Then, the convergence of $\mathcal{Q}_\tau^P(v_1,v_2)$
to a limit as $\tau\to\infty$ follows from the following facts:

(i) the quadratic form $\mathcal{Q}^{\nu}_\tau(\Psi,\Psi)$ converges to a limit
for any $\Psi\in\mathcal{S}$ (Theorem~\ref{l1}~(i));

(ii) $\mathcal{S}$ is dense in $\mathcal{H}(\kappa)$;

(iii) the quadratic forms $\mathcal{Q}^{\nu}_\tau(\Psi,\Psi)$, $\tau\in\mathbb{R}$,
are equicontinuous in $\mathcal{H}(\kappa)$ (Lemma~\ref{l3.7});

(iv) $R'_0T\Omega'\vec{v}\in\mathcal{H}(\kappa)$ for any $v\in\mathcal{P}$.\\
Hence, it remains to check the last fact. By (\ref{operatorOmega'}) and
(\ref{operatorT}),
$$
R'_0T\Omega'\vec{v}=R'_0\left(v+(\Gamma'\vec{v})^0
-\partial_t(\Gamma'\vec{v})^1\right).
$$
Due to (\ref{3.36}), we have
\begin{equation}\label{3.47}
\Vert \Gamma'\vec{v}(\cdot,t)\Vert_{\mathcal{H}(\kappa)}\equiv
\Vert (\Gamma'\vec{v})^0(\cdot,t)\Vert_{\ell^2(\kappa)}+
\Vert (\Gamma'\vec{v})^1(\cdot,t)\Vert_{\ell^2}
\le C\Vert v(\cdot,t)\Vert_{-\alpha}.
\end{equation}
Since $v+(\Gamma'\vec{v})^0\in L^1(\mathbb{R};\ell^2(\kappa))$,
then the bound~(\ref{3.26'})   gives
$$
\Vert R'_0\left(v+(\Gamma'\vec{v})^0\right)\Vert_{\mathcal{H}(\kappa)}
\le C \Vert v+(\Gamma'\vec{v})^0\Vert_{L^1(\mathbb{R};\ell^2(\kappa))}
\le C_1\Vert v\Vert_{L^1(\mathbb{R};\ell^2_{-\alpha})}.
$$
However, we can not apply  bound~(\ref{3.26'}) with  $(\Gamma'\vec{v})^1$ instead of $v$,
because $(\Gamma'\vec{v})^1(\cdot,t)\in \ell^2$ for any $t$ by (\ref{3.47}),
but $(\Gamma'\vec{v})^1(\cdot,t)\not\in \ell^2(\kappa)$, in general.
Now we study $R'_0 \dot{w}$ with $w:=(\Gamma'\vec{v})^1$.
Note first  that
$R'_0 v(y)=R'_\pm v(y)$ for $y\in\mathbb{Z}_\pm$ with
$$
(R'_\pm v)^j(y):=
\sum\limits_{x\in\mathbb{Z}_\pm}\int\limits_{-\infty}^{+\infty} G^{0j}_{t,\pm}(x,y)v(x,t)\,dt=
\sum\limits_{x\in\mathbb{Z}}\,\,\int\limits_{-\infty}^{+\infty} \mathcal{G}^{0j}_{t,\pm}(x-y)v_\pm(x,t)\,dt,
$$
where we use notation~(\ref{3.13}).
Then, in the Fourier transform,
$$
(\widehat{R'_\pm \dot w})^j(\theta)=
\int\limits_{-\infty}^{+\infty} \widehat{\mathcal{G}}^{0j}_{t,\pm}(\theta)\partial_t\widehat{w}_\pm(\theta,t)\,dt
=-\int\limits_{-\infty}^{+\infty} \widehat{\mathcal{G}}^{1j}_{t,\pm}(\theta)\widehat{w}_\pm(\theta,t)\,dt,
\qquad \theta\in\mathbb{T},
$$
by (\ref{hatcalG}). Hence, using the Parseval equality, we have
\begin{align*}
\Vert R'_\pm \dot w\Vert_{\mathcal{H}(\kappa)}
&\le C_1 \sum\limits_{j=0,1}\Vert (\widehat{R'_\pm\dot w})^j\Vert_{L^2(\mathbb{T})}
+C_2\Vert(\widehat{R'_\pm\dot w})^0\phi_\pm^{-1}\Vert_{L^2(\mathbb{T})}\\
&\le C
\int\limits_{-\infty}^{+\infty} \Vert \widehat{w}_\pm(\cdot,t)\Vert_{L^2(\mathbb{T})}\,dt\le
C_1\Vert w\Vert_{L^1(\mathbb{R};\ell^2)}.
\end{align*}
Therefore,
$$
\Vert R'_0\partial_t(\Gamma'\vec{v})^1\Vert_{\mathcal{H}(\kappa)}\le
\sum\limits_\pm \Vert R'_\pm \partial_t(\Gamma'\vec{v})^1\Vert_{\mathcal{H}(\kappa)}\le C
\Vert (\Gamma'\vec{v})^1\Vert_{L^1(\mathbb{R};\ell^2)}
\le C_1 \Vert v\Vert_{L^1(\mathbb{R};\ell^2_{-\alpha})}
$$
by (\ref{3.47}). Hence, $R'_0T\Omega'\vec{v}\in\mathcal{H}(\kappa)$.
This completes the proof of assertion~(i) of Theorem~\ref{tA}.
\medskip

Assertion~(ii) of Theorem~\ref{tA}  follows from the following lemma.
 \begin{lemma}\label{3.15}
 (1)
Let conditions $\mathbf{C}$, $\mathbf{S1}$, and $\mathbf{S2}$ hold.
Then the family of the measures $\{P_\tau,\,\tau\in \mathbb{R}\}$
 is weakly compact in the space $\mathfrak{C}^0_\beta$, with any  $\beta<\alpha<-3/2$,
and the bound holds:
\begin{equation}\label{p3.1}
\sup_{\tau\ge0} \int \br u\br^2_{\alpha,1,T} P_\tau(du) \le C(\alpha)<\infty,
\end{equation}
where the constant $C(\alpha)$ does not depend on $T>0$.

(2) Let conditions $\mathbf{C}$, $\mathbf{S1}$, $\mathbf{S3}$, and $\mathbf{S4}$ hold.
Then for every $v\in \mathcal{P}$, the characteristic functionals
  of $P_\tau$ converge   to a limit as $\tau\to\infty$,
\begin{equation}\label{gau}
\hat {P}_\tau (v)\equiv
\int e^{i[u,v]}P_\tau(du)\to \hat {P}_\infty (v),\quad \tau\to\infty.
\end{equation}
Here
$\hat {P}_\infty (v)=\hat {P}^{\nu}_\infty (T\Omega'\vec{v})$,
where $\hat {P}^{\nu}_\infty$ is  defined in (\ref{3.21}).
 \end{lemma}
{\bf Proof}\,
Similarly to (\ref{2.23-0}), we have
\begin{equation}\label{2.23}
P_\tau(\omega)=\mu_\tau(R^{-1}\omega)\quad
\mbox{ for any }\,\,\omega\in {\cal B}(\mathfrak{C}^1_\alpha)\quad \mbox{and }\,\,\tau>0,
\end{equation}
where $\mu_\tau$ is defined in Definition~\ref{def2.5}.
To prove the bound~(\ref{p3.1}), we apply (\ref{2.23}) and obtain
\begin{align*}
\int \br u\br^2_{\alpha,1,T} P_\tau(du)&=
\int \br RY\br^2_{\alpha,1,T} \,\mu_\tau(dY)=
\sup_{|s|\le T}\int \Vert U(s)Y\Vert^2_{\alpha}\, \mu_\tau(dY)\\
&
=\sup_{|s|\le T}\int \Vert Y\Vert^2_{\alpha}\, \mu_{s+\tau}(dY)
\le C(\alpha)<\infty
\end{align*}
by the bound~(\ref{2.18}).
The bound (\ref{p3.1}) and the Prokhorov theorem imply the weak compactness of the measures family
$\{P_\tau, \tau\in\mathbb{R}\}$ in the space $\mathfrak{C}^0_\beta$, $\beta<\alpha$.
This can be proved by a similar method as in the proof of Theorem~\ref{l2.1}~(iii).
\medskip

To prove (\ref{gau}), we use the inequality $\left|e^{i\xi}-1\right|\le |\xi|$ for $\xi\in\mathbb{R}$
and
bounds~(\ref{3.25}), (\ref{3.45'}) and obtain  that
$$
\left|\hat P_\tau(v)-\hat P^\nu_\tau(T\Omega'\vec{v})\right|
\le\mathbb{E}|\delta_\tau|\le \sqrt{\mathbb{E}(\delta^2_\tau)}\le
C\int\limits_{-\infty}^{+\infty}\langle t+\tau\rangle^{-1/2}\Vert v(\cdot,t)\Vert_{-\alpha}\,dt\to0,
\quad \tau\to\infty.
$$
It remains to apply Theorem~\ref{l2.1}~(iii)  and obtain that
\begin{equation}\label{3.43}
\hat P^\nu_\tau(T\Omega'\vec{v})\to\hat P^\nu_\infty(T\Omega'\vec{v}), \qquad \tau\to\infty.
\end{equation}
However, $T\Omega'\vec{v}\not\in \mathcal{P}$, in general.
 More precisely, convergence~(\ref{3.43})
follows from the following facts:

(i) the equality $\hat P^\nu_\tau(T\Omega'\vec{v})
=\hat{\nu}_\tau(R'_0T\Omega'\vec{v})$ holds by (\ref{2.23-0}) and (\ref{4.2});

(ii) $\hat{\nu}_\tau(\Psi)$ converges to a limit as $\tau\to\infty$
for any $\Psi\in\mathcal{S}$ (Theorem~\ref{l1}~(ii));

(iii) $\mathcal{S}$ is dense in $\mathcal{H}(\kappa)$ (evidently);

(iv) the characteristic functionals $\hat{\nu}_\tau(\Psi)$, $\tau\in\mathbb{R}$,
are equicontinuous in $\mathcal{H}(\kappa)$ (Lemma~\ref{l3.7});

(v) $R'_0T\Omega'\vec{v}\in\mathcal{H}(\kappa)$ for any $v\in\mathcal{P}$ (this was proved above).

Lemma~\ref{3.15} is proved.$\Box$
\medskip

This completes the proof of assertion~(ii) of Theorem~\ref{tA}.
Assertion~(iii) of Theorem~\ref{tA} is proved in next section.

\subsection{Mixing property of the limit measure $P_\infty$}\label{sec5}

 We first prove  the convergence~(\ref{2.33}) for the measure $P^\nu_\infty$.
 The invariance of  $P^\nu_\infty$
 w.r.t.  the group $S_\tau$, $\tau\in\mathbb{R}$,
follows from Theorem~\ref{l2.1}~(iii).
\begin{lemma}\label{l3.16}
The group $S_\tau$ is mixing w.r.t. the measure $P^\nu_\infty$, i.e.,
for any $f,g\in L^2(\mathfrak{C}^1_\alpha,P^\nu_\infty)$,
\begin{equation}\label{2.30}
\lim_{\tau\to\infty} \int f(S_\tau z)g(z)\,P^\nu_\infty(dz)
= \int f(z)\,P^\nu_\infty(dz) \int g(z)\,P^\nu_\infty(dz).
\end{equation}
In particular, the group $S_\tau$ is ergodic w.r.t. the measure $P^\nu_\infty$,
i.e.,
$$
\lim_{T\to\infty}\frac1{T}\int\limits_0^T f(S_\tau z)\,d\tau=\int f(z)\,P^\nu_\infty(dz)
 \quad (\mathrm{mod }\, P^\nu_\infty).
$$
\end{lemma}
{\bf Proof}\,
Since $P^\nu_\infty$ is a Gaussian measure with zero mean value,
it is enough to prove (see \cite{GS}) that for any $v_1,v_2\in\mathcal{P}$,
\begin{equation}\label{2.30+}
 I_\tau:= \int [S_\tau z,v_1][z,v_2]\,P^\nu_\infty(dz)\to0\quad \mbox{as }\,\, \tau\to\infty.
\end{equation}
Using (\ref{4.2}), (\ref{102}),  and (\ref{2.20}), we obtain
\begin{align}\label{3.55}
 I_\tau &= \int\left\langle Y,R'_0S_\tau^{-1} v_1\right\rangle
\left\langle Y,R'_0v_2\right\rangle\,\nu_\infty(dY)=
\left\langle Q^\nu_\infty(y_1,y_2),
(R'_0 S_\tau^{-1} v_1)(y_1)\otimes  (R'_0v_2)(y_2) \right\rangle\notag\\
&=\sum\limits_\pm\sum\limits_{x_1,x_2\in\mathbb{Z}_{\pm}}
\int\limits_{-\infty}^{+\infty}dt_1\int\limits_{-\infty}^{+\infty}
A_{\tau,\pm}(x_1,x_2,t_1,t_2)v_1(x_1,t_1)v_2(x_2,t_2)\,dt_2,
\end{align}
where, by definition,
$$
A_{\tau,\pm}(x_1,x_2,t_1,t_2):=
\sum\limits_{i,j=0,1}\sum\limits_{y_1,y_2\in\mathbb{Z}_\pm}
Q^{ij}_{\infty,\pm}(y_1,y_2) G^{0i}_{t_1+\tau,\pm}(x_1,y_1)G^{0j}_{t_2,\pm}(x_2,y_2).
$$
Similarly to (\ref{3.20}), we have
\begin{align}\label{2.31}
A_{\tau,\pm}(x_1,x_2,t_1,t_2)&:=\sum\limits_{i,j=0,1}\sum\limits_{y_1,y_2\in\mathbb{Z}}
q^{ij}_{\infty,\pm}(y_1-y_2) G^{0i}_{t_1+\tau,\pm}(x_1,y_1)G^{0j}_{t_2,\pm}(x_2,y_2)\notag\\
&=\frac{2}{\pi}\sum\limits_{i,j=0,1}
\int\limits_{\mathbb{T}}\hat q^{ij}_{\infty,\pm}(\theta)
\hat{\mathcal{G}}^{0i}_{t_1+\tau,\pm}(\theta) \hat{\mathcal{G}}^{0j}_{t_2,\pm}(\theta)
\sin(x_1\theta)\sin(x_2\theta)\,d\theta\notag\\
&=\frac{2}{\pi}
\int\limits_{\mathbb{T}}
\cos\left(\phi_\pm(\theta)(t_1+\tau-t_2)\right)\hat q^{00}_{\infty,\pm}(\theta)
\sin(x_1\theta)\sin(x_2\theta)\,d\theta.
\end{align}
Hence, applying Lemma~\ref{rem2.5}~(ii), formulas~(\ref{qinfty}),
 and Fei\'er's theorem (if $\kappa_\pm=0$), we obtain
\begin{equation}\label{3.54}
\left|A_{\tau,\pm}(x_1,x_2,t_1,t_2)\right|\le C\int\limits_{\mathbb{T}}\left|\hat q^{00}_{\infty,\pm}(\theta)
\sin(x_1\theta)\sin(x_2\theta)\right|\,d\theta\le C_1+C_2(|x_1|+|x_2|),
\end{equation}
where the constants $C_1$ and $C_2$ do not depend on $x_1,x_2\in\mathbb{Z}_\pm$ and $C_2=0$ if $\kappa_\pm\not=0$.
Since $v_1,v_2\in\mathcal{P}$,
it follows from (\ref{3.55}) and (\ref{3.54}) that
to prove (\ref{2.30+}) it suffices to check the  convergence
\begin{equation}\label{3.45}
A_{\tau,\pm}(x_1,x_2,t_1,t_2)\to0\qquad\mbox{as }\,\,\tau\to\infty,
\end{equation}
for fixed values of $x_1,x_2\in\mathbb{Z}_\pm\setminus\{0\}$ and $t_1,t_2\in\mathbb{R}$.
We denote by $R_{\tau,\pm}$ the integrand in the r.h.s. of (\ref{2.31}) and rewrite it in the form
\begin{align*}
R_{\tau,\pm}(\theta)&\equiv R_{\tau,\pm}(\theta;x_1,x_2,t_1,t_2)
=\cos\left(\phi_\pm(\theta)(t_1+\tau-t_2)\right)
\hat q^{00}_{\infty,\pm}(\theta)\sin(x_1\theta)\sin(x_2\theta)\\
&=\cos\left(\phi_\pm(\theta)\tau\right)a_\pm(\theta)+
\sin\left(\phi_\pm(\theta)\tau\right)b_\pm(\theta),
\end{align*}
where
\begin{align*}
a_\pm(\theta)&\equiv a_\pm(\theta;x_1,x_2,t_1,t_2)=\cos\left(\phi_\pm(\theta)(t_1-t_2)\right)
\hat q^{00}_{\infty,\pm}(\theta)\sin(x_1\theta)\sin(x_2\theta),\\
b_\pm(\theta)&\equiv b_\pm(\theta;x_1,x_2,t_1,t_2)=-\sin\left(\phi_\pm(\theta)(t_1-t_2)\right)
\hat q^{00}_{\infty,\pm}(\theta)\sin(x_1\theta)\sin(x_2\theta),
\end{align*}
and $a_\pm,b_\pm\in L^1(\mathbb{T})$.
Choose a $\delta> 0$ and introduce a partition of unity,
$f(\theta)+g(\theta)=1$, where $f$ and $g$ are nonnegative functions in $C^\infty(\mathbb{T})$,
$\mathrm{supp } f \subset \mathcal{O}_\delta(0)$,
 $\mathrm{supp }\,g\cap \mathcal{O}_{\delta/2}(0)=\emptyset$.
 We split $A_{\tau,\pm}$ into the sum of two integrals,
 $$
A_{\tau,\pm}= \frac{2}{\pi}\int\limits_{\mathbb{T}} f(\theta) R_{\tau,\pm}(\theta)\,d\theta
+\frac{2}{\pi}\int\limits_{\mathbb{T}} g(\theta) R_{\tau,\pm}(\theta)\,d\theta
\equiv A_{\tau,\pm}^f+A_{\tau,\pm}^g.
$$
On the one hand, $\forall\varepsilon>0$  $\exists\delta>0$
such that $|A_{\tau,\pm}^f|\le C\varepsilon$ uniformly in $\tau$.
On the other hand, the phase functions $\phi_\pm(\theta)$ are smooth
and $\phi'_\pm(\theta)\not= 0$ on the support of the $g$.
 Hence, the oscillatory integrals in $A_{\tau,\pm}^g$ vanish by the Lebesgue--Riemann Theorem.
 Therefore, the convergence~(\ref{3.45}) holds and
  Lemma~\ref{l3.16} is proved.
 \medskip

 Now we check assertion~(\ref{2.33}) for the limit measure $P_\infty$.
 We prove that for any $v_1,v_2\in\mathcal{P}$
 \begin{equation}\label{3.46}
 I'_\tau:= \int [S_\tau u,v_1][u,v_2]\,P_\infty(du)\to0
 \qquad \mbox{as }\,\, \tau\to\infty.
\end{equation}
Applying (\ref{2.9}) gives
$$
 I'_\tau= \int[z,T\Omega'\vec{v}_1(\cdot,t-\tau)]\,
 [z,T\Omega'\vec{v}_2]\,P^\nu_\infty(dz)
 = \int[S_\tau z,T\Omega'\vec{v}_1]\, [z,T\Omega'\vec{v}_2]\,P^\nu_\infty(dz).
$$
Put $f_i(z):=[z,T\Omega'\vec{v}_i]$,
$i=1,2$. Then, $f_i\in L^2(\mathfrak{C}^1_\alpha,P^\nu_\infty)$, since
$$
\int|f_i(z)|^2\,P^\nu_\infty(dz)=
\int[u,v_i]^2\,P_\infty(du)<\infty
$$
by (\ref{2.9}).
Therefore, we apply (\ref{2.30}) and obtain
$$
 I'_\tau= \int f_1(S_\tau z)f_2(z)\,P^\nu_\infty(dz)
\to \int f_1(z)\,P^\nu_\infty(dz) \int f_2(z)\,P^\nu_\infty(dz)
\qquad\mbox{as }\,\,\tau\to\infty.
$$
Finally,
$$
\int f_i(z)\,P^\nu_\infty(dz)=\int [u,v_i]\,P_\infty(du)=0,
$$
because $P_\infty$ has zero mean value.
Hence, (\ref{3.46}) holds. This completes the proof of
Theorem~\ref{tA}.
$\Box$

\appendix

\setcounter{section}{1}
\setcounter{theorem}{0}
\setcounter{equation}{0}

\section*{Appendix: Homogeneous harmonic chain}

Let condition~(\ref{0.1}) hold. Then the problem~(\ref{1})--(\ref{4}) becomes
\begin{equation*}\label{a.0}
\left\{\begin{array}{ll}
 \ddot u(x,t)=(\nu^2\Delta_L-\kappa^2)u(x,t),& x\in\mathbb{Z},\quad t>0,
\\
u(x,0)=u_0(x),\quad \dot u(x,0)=v_0(x),& x\in\mathbb{Z}.
\end{array}
\right.
\end{equation*}
At first, we state results concerning the statistical solutions $\mu_t$, $t\in\mathbb{R}$.
\begin{lemma}\label{t4} (see \cite[Theorem~A]{DKM1})
 Let $\alpha<-1/2$ and condition~(\ref{0.1}) hold.
 Then all assertions of Theorem~\ref{t2.11} hold, where
 the quadratic form ${\cal Q}_\infty$  has the matrix kernel   \begin{equation}\label{4.5}
Q_\infty(x,y)=q_\infty(x-y),
\end{equation}
 $q_\infty(x)=F^{-1}_{\theta\to x}[\hat q_\infty(\theta)]$
and $\hat q_\infty(\theta)=\left(\hat q^{ij}_\infty(\theta)\right)$ is defined in (\ref{2.8}).
\end{lemma}
{\bf Proof of Theorem~\ref{tB}}.
Introduce the adjoint operator $R'$  to the operator  $R$ defined in (\ref{R}),
\begin{equation}\label{scpro}
[RY,v]=\langle Y,R'v\rangle
 \quad\mbox{for }\,\, Y\in{\cal H}_\alpha\quad
 \mbox{and }\,\, v\in \mathcal{P}.
\end{equation}
Then for $v\in\mathcal{P}$,
\begin{equation}\label{4.4}
(R'v)(x)=\Big(\sum_{y\in\mathbb{Z}}\int\limits_{-\infty}^{+\infty} \mathcal{G}^{00}_t(y-x)v(y,t)dt,
\sum_{y\in\mathbb{Z}}\int\limits_{-\infty}^{+\infty} \mathcal{G}^{01}_t(y-x)v(y,t)dt\Big),
\end{equation}
where $\mathcal{G}^{ij}_t$ is defined as $\mathcal{G}^{ij}_{t,+}$ in (\ref{3.2}) and (\ref{hatcalG})
but with $\phi(\theta)$ instead of $\phi_+(\theta)$.
It follows from (\ref{scpro}) and (\ref{2.23})
that for $v_1,v_2\in\mathcal{P}$,
\begin{equation*}
\begin{split}
\mathcal{Q}^P_\tau(v_1,v_2)&:=\int[u,v_1][u,v_2]\,P_\tau(du)=
\int[RY,v_1][RY,v_2]\,\mu_\tau(dY)\\
&=\int\langle Y,R'v_1\rangle\langle Y,R'v_2\rangle\,\mu_\tau(dY)=\langle Q_\tau(x,y),R'v_1(x)\otimes R'v_2(y)\rangle.
\end{split}
\end{equation*}
Since $\kappa\not=0$, then
$R'v\in{\cal S}$ for any $v\in\mathcal{P}$.
 Hence, we can apply Lemma~\ref{t4} and obtain
$$
\mathcal{Q}^P_\tau(v_1,v_2)\to\langle Q_\infty(x,y),R'v_1(x)\otimes R'v_2(y)\rangle
\quad \mbox{as }\,\, \tau\to\infty.
$$
Hence,
$\mathcal{Q}^P_\infty(v_1,v_2)=\langle Q_\infty(x,y),R'v_1(x)\otimes R'v_2(y)\rangle$.
Now we check formula~(\ref{2.21}).
Using (\ref{4.5}) and (\ref{4.4}), we have
$$
Q_\infty^P(x_1,x_2,t_1,t_2)=Q_*^P(x_1-x_2,t_1,t_2),
$$
where in the Fourier transform $x\to\theta$
$$
\hat Q_*^P(\theta,t_1,t_2)=\sum\limits_{x\in\mathbb{Z}}e^{i\theta x}Q_*^P(x,t_1,t_2)=
\sum\limits_{i,j=0}^1 \hat{\mathcal{G}}_{t_1}^{0i}(\theta)\hat q^{ij}_\infty(\theta)\hat {\mathcal{G}}_{t_2}^{0j}(\theta),
\quad \theta\in\mathbb{T},\quad t_1,t_2\in\mathbb{R}.
$$
Using formulas
$\hat q^{11}_{\infty}(\theta)=\phi^2(\theta)\hat q^{00}_{\infty}(\theta)$,
$\hat q^{10}_{\infty}(\theta)=-\hat q^{01}_{\infty}(\theta)$, and
 (\ref{hatcalG}) with $\phi_+\equiv\phi$, we have
\begin{align*}
\hat Q_*^P(\theta,t_1,t_2)=&\cos\left(\phi t_1\right)\hat q^{00}_\infty(\theta)\cos\left(\phi t_2\right)-
\sin\left(\phi t_1\right)\phi^{-1}\hat q^{01}_\infty(\theta)\cos\left(\phi t_2\right)\\
&+
\cos\left(\phi t_1\right)\hat q^{01}_\infty(\theta)\phi^{-1}\sin\left(\phi t_2\right)
+\sin\left(\phi t_1\right)\hat q^{00}_\infty(\theta)\sin\left(\phi t_2\right)\\
=&\cos\left(\phi (t_1-t_2)\right)\hat q^{00}_\infty(\theta)-
\sin\left(\phi (t_1-t_2)\right)\phi^{-1}\hat q^{01}_\infty(\theta)
\end{align*}
with $\phi\equiv \phi(\theta)$.
This implies (\ref{2.21}).
\smallskip

Now we prove the convergence~(\ref{1.8i}) by a similar way as in Theorem~\ref{tA}.
Assertion~(\ref{1.8i}) follows from the bound~(\ref{p3.1}) and
convergence~(\ref{gau}).
The bound~(\ref{p3.1}) can be proved in the same way as in  Theorem~\ref{tA}.
Lemma~\ref{t4} implies that for any $\Psi\in{\cal S}$,
$$
\hat\mu_t(\Psi)\to\hat\mu_\infty(\Psi)\quad \mbox{as }\,\, t\to\infty.
$$
Using (\ref{scpro}) and taking $\Psi:=R'v$, we obtain
$$
\hat {P}_\tau (v)=
\int e^{i[u,v]}P_\tau(du)\equiv
\int e^{i\langle Y,R'v\rangle} \mu_\tau(dY)
\to\exp\left\{-\frac12 {\cal Q}_\infty(R'v,R'v)\right\},\quad \tau\to\infty,
$$
where the quadratic form $\mathcal{Q}_\infty$ is introduced in Lemma~\ref{t4}.
Theorem~\ref{tB} is proved.
$\Box$
\medskip

Now we verify the mixing property~(\ref{2.33}) for the limit measure $P_\infty$.
 The invariance of the measure $P_\infty$
 w.r.t.  the shifts in time and in space
follows from convergence~(\ref{1.8i}) and (\ref{4.5}).
Since the measure $P_\infty$ is Gaussian with zero mean value,
it is enough to prove that for any $v_1,v_2\in\mathcal{P}$,
\begin{equation}\label{2.30h}
 I_\tau:= \mathbb{E}_\infty \left([S_\tau u,v_1][u,v_2]\right)\to0\qquad \mbox{as }\,\, \tau\to\infty,
\end{equation}
where $\mathbb{E}_\infty$ denotes the integral w.r.t. the limit measure $P_\infty$.

Indeed, using  (\ref{scpro}) and (\ref{4.4}), we obtain
\begin{align*}
 I_\tau &= \int\left\langle Y,R'S_\tau^{-1} v_1\right\rangle
\left\langle Y,R'v_2\right\rangle\,\mu_\infty(dY)
=
\frac{1}{2\pi}
\int\limits_{\mathbb{T}}\left( \hat q_\infty(\theta),
\widehat{(R'S_\tau^{-1} v_1)}(\theta)\otimes \widehat{ (R'v_2)}(\theta) \right)\,d\theta\\
&
=\frac{1}{2\pi}\int\limits_{\mathbb{T}}d\theta
\int\limits_{-\infty}^{+\infty}dt_1
\int\limits_{-\infty}^{+\infty}
 B_\tau(t_1,t_2,\theta)
\overline{\hat v_1(\theta,t_1)} \hat{v}_2(\theta,t_2)\,dt_2,
\end{align*}
where
 the function $B_\tau(t_1,t_2,\theta)$ is of the form
\begin{align}
B_\tau(t_1,t_2,\theta)&:=\sum\limits_{i,j=0}^1
\hat{\mathcal{G}}_{t_1+\tau}^{0i}(\theta)\hat q^{ij}_\infty(\theta)\hat {\mathcal{G}}_{t_2}^{0j}(\theta)
\notag\\
&=
\cos\left(\phi(\theta) (t_1+\tau-t_2)\right)\hat q^{00}_\infty(\theta)-
\sin\left(\phi(\theta) (t_1+\tau-t_2)\right)\phi^{-1}(\theta)\hat q^{01}_\infty(\theta). \notag
\end{align}
We represent $I_\tau$ as follows:
\begin{equation}\label{2.32h}
I_\tau=\sum\limits_{\pm}\frac{1}{2\pi}\int\limits_{\mathbb{T}}
e^{\pm i\phi(\theta)\tau}c_\pm(\theta)\,d\theta,
\end{equation}
where
$$
c_\pm(\theta):=\frac12\int\limits_{-\infty}^{+\infty}dt_1
\int\limits_{-\infty}^{+\infty}
  e^{\pm i\phi(\theta)(t_1-t_2)}\left(\hat q^{00}_\infty(\theta)
  \pm i\phi^{-1}(\theta)\hat q^{01}_\infty(\theta)\right)
\overline{\hat v_1(\theta,t_1) }\hat{v}_2(\theta,t_2)\,dt_2.
$$
 Note that $c_\pm(\theta)\in L^1(\mathbb{T})$ by Lemma~\ref{rem2.5}~(ii) and formulas~(\ref{qinfty})
 with $\phi_\pm\equiv\phi$.
 Hence,
 the oscillatory integrals in (\ref{2.32h})  vanish by  the Lebesgue--Riemann Theorem.
 Therefore, the assertions~(\ref{2.30h}) and (\ref{2.33}) hold.
\medskip

Similarly to~(\ref{2.33}), we can  check the following assertion.
\begin{lemma}
Let $\mathcal{S}_h$, $h\in\mathbb{Z}$, denote the shifts in space,
$\mathcal{S}_h u(x,t)=u(x+h,t)$. Then,
for any $f,g\in L^2(\mathfrak{C}^1_\alpha,P_\infty)$,
$$ 
\lim_{h\to\infty} \mathbb{E}_\infty f(\mathcal{S}_h u)g(u)=\mathbb{E}_\infty f\,
\mathbb{E}_\infty g.
$$
\end{lemma}
{\bf Proof}\,
Indeed, it suffices to check that
$I_h:= \mathbb{E}_\infty \left([\mathcal{S}_h u,v_1][u,v_2]\right)\to0$ as $h\to\infty$.
Using  (\ref{scpro}) and (\ref{4.4}), we obtain
$$ 
 I_h = \int\langle Y,R' \mathcal{S}_h^{-1} v_1\rangle
\langle Y,R'v_2\rangle\,\mu_\infty(dY)
=\frac1{2\pi}\int\limits_{\mathbb{T}} e^{ih\theta}D(\theta)\,d\theta,
$$
where
$$
D(\theta):=\int\limits_{-\infty}^{+\infty}dt_1
\int\limits_{-\infty}^{+\infty}
 \left(
\cos\left(\phi (t_1-t_2)\right)\hat q^{00}_\infty(\theta)-
\sin\left(\phi (t_1-t_2)\right)\phi^{-1}\hat q^{01}_\infty(\theta)\right)
\overline{\hat v_1(\theta,t_1)} \hat{v}_2(\theta,t_2)\,dt_2.
$$
Therefore, $I_h$ vanishes as $h\to\infty$
by the Lebesgue--Riemann Theorem, because $D(\theta)\in L^1(\mathbb{T})$.
$\Box$

The following lemma generalizes the convergence~(\ref{2.33}).
\begin{lemma}
The group $S_\tau$ is mixing of order $r\ge1$ w.r.t. the measure $P_\infty$,  i.e.,
  for any $f_0,\ldots,f_r\in L^{r+1}(\mathfrak{C}^1_\alpha,P_\infty)$,
$$ 
\lim_{\tau_1,\ldots,\tau_r\to\infty} \int
 f_0(u)f_1(S_{\tau_1} u)\cdot\ldots\cdot f_r\left(S_{\tau_1+\ldots+\tau_r} u\right)\,P_\infty(du)
 =\prod_{i=0}^{r}\int f_i(u)P_\infty(du).
$$
\end{lemma}
{\bf Proof}\,
Since the measure $P_\infty$ is Gaussian with zero mean value,
it is enough to prove that for any $v_0,\ldots,v_r\in\mathcal{P}$,
\begin{equation}\label{2.36}
 I_{\tau_1,\ldots,\tau_r}
 :=\mathbb{E}_\infty \left([u,v_0][S_{\tau_1} u,v_1]\cdot\ldots\cdot[S_{\tau_1+\ldots+\tau_r}u,v_r]\right)
 \to0  \qquad \mbox{as }\,\, \tau\to\infty.
\end{equation}
At first, note that (see \cite[Ch.III, \S\,1]{GS})
$$
\mathbb{E}_\infty \left([u,v_1]\cdot\ldots\cdot[u,v_n]\right)=
\left\{
\begin{array}{ll}
0, & \text{if $n$ is odd,}\\
\sum\prod
\mathbb{E}_\infty \left([u,v_i][u,v_j]\right), &
 \text{if $n$ is even.}
\end{array}
\right.
$$
Here the sum is taken over all partitions of $\{v_1,\ldots,v_n\}$ into pairs,
the product is taken over all pairs of the partition
(the pairs that differ by the permutation of elements are considered as one).
For example, if $n=4$, there are three partitions of $\{v_1,v_2,v_3,v_4\}$ into pairs and
$$
\mathbb{E}_\infty \left([u,v_1]\cdot\ldots\cdot[u,v_4]\right)=
b_{12}b_{34}+b_{13}b_{24}+b_{14}b_{23},
\quad \mbox{where }\,b_{ij}:=\mathbb{E}_\infty \left([u,v_i][u,v_j]\right).
$$
Hence,
$ I_{\tau_1,\ldots,\tau_r}=0$ if $r$ is even.
If $r$ is odd, then
\begin{align}\label{2.38}
I_{\tau_1,\ldots,\tau_r}&=
\mathbb{E}_\infty \left([u,v_0][u,S_{\tau_1}^{-1} v_1]\cdot\ldots\cdot[u,S_{\tau_1+\ldots+\tau_r}^{-1}v_r]\right)\notag\\
&=\sum_{k=1}^r\mathbb{E}_\infty \left([u,v_0][u,S_{\tau_1+\ldots+\tau_k}^{-1}v_k]\right)
\cdot\left(\sum\prod B_{ij}\right),
\end{align}
where the inner sum is taken over all partitions of $\{v_1,\ldots,v_{k-1},v_{k+1},\ldots,v_r\}$ into pairs,
the product is taken over all pairs of the partition,
and
$$
B_{ij}:=\mathbb{E}_\infty \left(\big[u,S_{\tau_1+\ldots+\tau_i}^{-1}v_i\big]
\big[u,S_{\tau_1+\ldots+\tau_j}^{-1}v_j\big]\right),
\qquad i,j=1,\ldots,k-1,k+1,\dots,r.
$$
Since the measure $P_\infty$ is invariant w.r.t. $S_\tau$, 
$$ 
B_{ij}\le\sqrt{\mathbb{E}_\infty([u,v_i]^2)}\sqrt{\mathbb{E}_\infty([u,v_j]^2)}\le C<\infty.
$$
Furthermore, (\ref{2.30h}) implies that for any $k\ge1$
\begin{equation}\label{2.40}
\mathbb{E}_\infty \left([u,v_0][u,S_{\tau_1+\ldots+\tau_k}^{-1}v_k]\right)\to0
\qquad\mbox{as }\,\tau_1,\ldots,\tau_k\to+\infty.
\end{equation}
Formulas (\ref{2.38})--(\ref{2.40}) imply the convergence~(\ref{2.36}).
$\Box$

\medskip\medskip

{\bf Acknowledgment}.
This work was supported by the Russian Science  Foundation (Grant no. 19-71-30004).


\end{document}